\documentclass[aps,groupedaddress,showpacs]{revtex4}
\usepackage{epsfig,graphicx}
\usepackage{amsmath,amssymb,amsthm}

\newtheorem{theorem}{Theorem}
\newtheorem{defn}{Definition}
\newtheorem{obs}{Observation}
\newtheorem{lemma}{Lemma}
\newtheorem{cor}{Corollary}[lemma]
\newtheorem{rem}{Remark}

\input epsf

\begin{document}

\title[Confinement of unstretchable manifolds] {Limitations on the
  smooth confinement of an unstretchable manifold}

\author{S. C. Venkataramani} \email{shankar@math.uchicago.edu}
\affiliation{Dept. of Mathematics and MRSEC\\ University of Chicago,
Chicago, IL 60637}

\author{T. A. Witten}
\affiliation{James Franck Inst., Dept. of Physics and MRSEC \\ University
    of Chicago, Chicago, IL 60637}  

\author{E. M. Kramer} 
\affiliation{Physics Dept.\\ Simon`s Rock
College, 84 Alford Road., Great Barrington, MA 01230}

\author{R. P.  Geroch} \affiliation{Enrico Fermi Inst. and Dept. of
Physics\\ University of Chicago, Chicago, IL 60637}

\date{January 25, 2000}

\begin{abstract}
   We prove that an $m$-dimensional unit ball $D^m$ in the Euclidean
  space ${\mathbb R}^m$ cannot be isometrically embedded into a
  higher-dimensional Euclidean ball $B_r^d \subset {\mathbb R}^d$ of
  radius $r < 1/2$ unless one of two conditions is met --
\begin{enumerate} 
\item  The embedding manifold
  has dimension $d \geq 2m$.  
\item The embedding is not smooth.
\end{enumerate}
The proof uses differential geometry to show that if $d<2m$ and the
embedding is smooth and isometric, we can construct a line from the
center of $D^m$ to the boundary that is geodesic in both $D^m$ and in
the embedding manifold ${\mathbb R}^d$.  Since such a line has length
1, the diameter of the embedding ball must exceed 1.
\end{abstract}

\pacs{02.40.-k, 02.40.Ky, 02.40.Ma, 02.40.Sf}

\maketitle

\section{Introduction} \label{sec:intro}

Mechanical deformation of objects is a part of our everyday
experience.  Objects like coil springs, wrapping films or blood cells
are designed to be deformed in specified ways.  Clearly the shape of
an object influences how it may be deformed. We will be interested in
looking at how the intrinsic geometry of an object restricts the ways
in which the object can be deformed.

The mathematical notion of a Riemannian manifold gives a way of
describing the shape of an object and its deformation.  A Riemannian
manifold is a set that has the topological structure of an Euclidean
space ${\mathbb R}^n$ locally and is equipped with a local measure of
distance, called the metric. An interesting question about
two-dimensional or one-dimensional objects (manifolds) in three
dimensional space is whether they can be deformed ``isometrically''---
that is, without changing lengths of lines in the object. Since an
isometric deformation of a material causes no stretching, it is of
practical importance to know what kinds of deformation can be made
isometrically.

A classical problem in differential geometry is the study of isometric
embeddings of a flat 2-dimensional sheet in ${\mathbb R}^3$. This is
the study of {\it developable
surfaces}~\cite{DiffGeom1,DiffGeom2}. This problem has seen a recent
resurgence of interest \cite{Pomeau,cones_exp,cones_thry,maha,Basile}
because of its connections with the nature of the singularities in a
crumpled sheet
\cite{buckling,buckling2,acoustic,cornell,Li-Witten,wit.sci}.

Differential geometers are interested in the much more general
question of whether a given manifold can be embedded in another
manifold isometrically. A sphere and a M\"{o}bius strip are examples
of $2$ dimensional manifolds that cannot be isometrically embedded in
${\mathbb R}^2$ but can be isometrically embedded in ${\mathbb R}^3$.
There is a fundamental theorem, due to Nash~\cite{nash} showing that
every manifold can be embedded in an Euclidean Space ${\mathbb R}^d$
of a sufficiently large dimension.  However, the smallest dimension
which allows this embedding depends on the additional structure
imposed on the manifold and the embedding, and is the subject of
current research. A more complete discussion of this question and
improvements of Nash's results can be found in
references~\cite{GR,Gro}.

In this paper we study the {\it confinability} of an embedded
manifold.  We will say that a $m$ dimensional object is {\em
confinable} in $d$ dimensions if it is possible to smoothly deform the
object without inducing any stretching so that it lies inside an
arbitrarily small $d$ dimensional sphere. Of course, for this to make
sense, we need $d > m$. Note that $d > m$ does not imply that such
deformations are always possible.  A two dimensional sheet confined in
a shrinking sphere develops singularities -- a phenomenon described
informally as {\em crumpling}.  Thus there is a lower bound on the
size of the enclosing sphere which can contain the sheet smoothly and
isometrically, that is without any stretching.  We show below that
this bound exists whenever the embedding space has fewer than twice
the dimensions of the embedded manifold: the greatest distance between
two embedded points, or the ``span'' of the embedded sheet, must be at
least half the intrinsic diameter of the sheet. The converse of our
theorem states that an $m$-dimensional manifold may be confined into
an arbitrarily small sphere in ${\mathbb R}^{2m}$.  This converse may
be readily shown by an explicit construction \cite{Kramer.PRL}.

The paper is organized as follows -- We begin by discussing the
problem in Section.~\ref{sec:problem}. In this section, we also
illustrate the ideas behind our proof using the case of a two-sheet
embedded in 3 dimensions. Section~\ref{sec:rvw} reviews the theorems
of differential geometry on which our theorem is based, and
Section~\ref{sec:proofs} contains the proof of the theorem, together
with the proofs of seven lemmas used in the proof of the
theorem. Section~\ref{sec:conclusions} discusses implications of the
theorem, related ideas, and possible generalizations.

In the body of the paper, we use many standard definitions and results
from differential geometry. For completeness, we include an appendix,
where we discuss some of these definitions and results. In this
appendix, we have appended brief remarks to the standard definitions
of the various mathematical objects to give some physical intuition
about some of these notions.  A complete discussion of these and
related topics can be found in~\cite{manifolds,waldbook}.

\section{Constrained Isometric immersions} \label{sec:problem}

Our study concerns the distortions of an object in space.
Accordingly, we must characterize mappings of a manifold $M$
representing the object into another manifold $\tilde M$ representing
the space. Let $\phi:M \rightarrow \tilde{M}$ be a smooth mapping and
let $\phi_* :T_p M \rightarrow T_{\phi(p)}\tilde{M}$ be the induced
map between the tangent space of $M$ at $p$ and the tangent space of
$\tilde{M}$ at $\phi(p)$.

\begin{defn} $\phi$ is an {\em immersion} if $\phi_* : T_p M \rightarrow
  T_{\phi(p)} \tilde{M}$ is one-to-one for every $p \in M$.
\end{defn}

If $M$ and $\tilde{M}$ are Riemannian manifolds and $\phi$ is an
immersion, we will say that $\phi$ is an {\em isometric immersion} if 
$$
\langle X, Y \rangle_p = \langle\phi_*X,\phi_*Y\rangle_{\phi(p)},
$$ for all $p \in M$, all $X,Y \in T_pM$. 

We are now in a position to state our problem precisely :

Let $D^m = \{{\mathbf p} \in {\mathbb R}^m \,\, | \,\, \|p\| \leq 1\}$
be the closed unit disk in ${\mathbb R}^m$ and $\bar{B}_r = \{{\mathbf
q} \in {\mathbb R}^d \,\, | \,\, \|q\| \leq r\}$ be the closed ball
with radius $r$ in ${\mathbb R}^d$. Given an $r > 0$, does there exist
an isometric immersion $\phi : D^m \rightarrow \bar{B}_r$ ? By analogy
with the case of a $2$-sheet, we will call the isometric immersion
$\phi : D^m \rightarrow \bar{B}_r$, if it exists, a smooth {\it
confinement} of an $m$-sheet. Such a smooth confinement is always
possible when $d \geq 2m$ \cite{Kramer.PRL}.  An explicit realization
is given in Section~\ref{sec:proofs}.

In Theorem~\ref{thm:crumple}, we show that, if $d < 2m$, we can choose
$r$ sufficiently small so that there is no such immersion.  We will
prove this theorem in Sec.~\ref{sec:proofs}. In the rest of this
section, we discuss the idea behind our proof of the theorem using the
example of a 2 dimensional surface in three dimensions.

The isometric embedding of 2-dimensional ``flat'' sheets in ${\mathbb
R}^3$ is a problem that has been explored by classical geometers for
more than a century. This is the study of {\em developable surfaces}
in three dimensional space~\cite{DiffGeom2}.  A surface is developable
if it has zero intrinsic curvature everywhere (with the exception of
possible singular points and lines).  Developable surfaces are
remarkable because, as proved by the {\it Theorema Egregium}
\cite{DiffGeom2}, they may be constructed by deforming a portion of
the plane without stretching it.  A thin sheet of paper is essentially
unstretchable, and the allowed deformations of the sheet provide the
archetypal model of a developable surface. A sheet may be smoothly
bent into a portion of the surface of a cone or cylinder, but not into
a portion of the surface of a sphere. The former are developable while
the latter is not.

\begin{figure}[tbh]
\epsfxsize = 0.7 \hsize
\centerline{\epsfbox{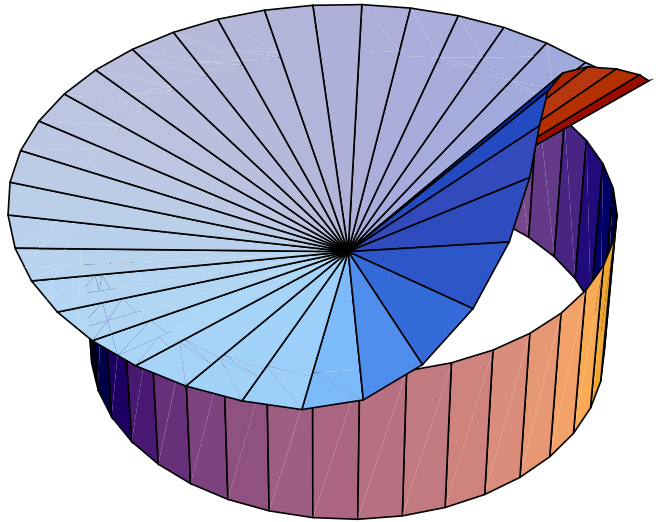}}
\vspace{0.0cm}
\caption{ A Cone and a cylinder that are made by smooth deformations
of a portion of a sheet without stretching. Both the cone and the
cylinder are developable surfaces and the straight lines shown are
generators for the surfaces. Note that the Cone has a singularity at
its apex, where it has an infinite curvature.  }
\label{fig:develop}       
\end{figure}

Consider a flat sheet smoothly bent into a portion of a cone or a
cylinder without stretching as in Fig~\ref{fig:develop}.  Through each
point on the surface, there exists at least one line, the {\em
generator}, that is a geodesic (straight line) both in the sheet when
it is flattened out in $\mathbb{R}^2$ and in the sheet as it is
embedded in $\mathbb{R}^3$. These generators can be characterized by
the fact that $p$ and $q$ are two points on a generator if and only if
$d(p,q)$, the distance between $p$ and $q$ on the sheet is the same as
$r(p,q)$, the distance between the images $\phi(p)$ and $\phi(q)$ in
the embedding space $\mathbb{R}^3$.  Thus, in these examples, given a
point $p$ in the sheet, one can find a straight line that extends from
$p$ to some point $q$ on the boundary. The length of this line in
$\mathbb{R}^3$ is equal to the distance between $p$ and $q$ in the
sheet. Consequently, no sphere whose diameter is smaller than half the
diameter of the sheet can contain the sheet.

We now consider a general isometric embedding of a $2$-disk in
${\mathbb R}^3$. We can choose global Cartesian co-ordinates on the
disk and also on ${\mathbb R}^3$. In these co-ordinates, the embedding
is given by three real-valued function $r^a(x^{\alpha}), a = 1,2,3$,
where $x^{\alpha},\alpha = 1,2$ are Cartesian co-ordinates on the
disk. If we let ${\mathbf r}(x^{\alpha})$ denote a vector valued
function with components $r^a(x^{\alpha})$, the {\em strain} is given
by \cite{landau}
\begin{equation}
u_{\alpha \beta} = \frac{\partial {\mathbf r}}{\partial x^{\alpha}} \cdot
\frac{\partial {\mathbf r}}{\partial x^{\alpha}} - \delta_{\alpha \beta}
\label{eq:strain}
\end{equation}
and the curvature is given by
\begin{equation}
C_{\alpha \beta} = \frac{\partial^2 {\mathbf r}}{\partial
x^{\alpha}\partial x^{\beta}} \cdot {\mathbf n},
\label{eq:extr_curv}
\end{equation}
where $\mathbf{n}$ is the unit normal to the surface. The {\em
principal curvatures} are the eigenvalues of the symmetric matrix
$C_{\alpha \beta}$. The {\em Gaussian curvature} is defined as the
product of the principal curvatures.

To show the existence of generators for all isometric immersions of a
flat sheet in $\mathbb{R}^3$, we need Gauss' {\em Theorema Egregium}
\cite{DiffGeom2}. This theorem asserts that the isometric immersion
yields a developable surface, so that the Gaussian curvature is zero
everywhere. Consequently, at least one of the principal curvatures is
zero at every point on the surface.

Let $p$ denote the center of the 2-disk that is embedded in ${\mathbb
R}^3$. We first consider the case where only one principal curvature
is zero at $p$. Since only one principal curvature is zero, there is a
unique generator through $p$. It can be shown that this generator can
be extended until it runs into the boundary of the disk at $q$. Since
the image of the generator $pq$ in $\mathbb{R}^3$ is also a straight
line, $d(p,q) = 1$ implies that $r(p,q) = 1$. Consequently, the sheet
cannot be embedded inside a shell with a diameter less than $1$.

\begin{figure}[tbh]
  \epsfxsize = 0.7 \hsize \centerline{\epsfbox{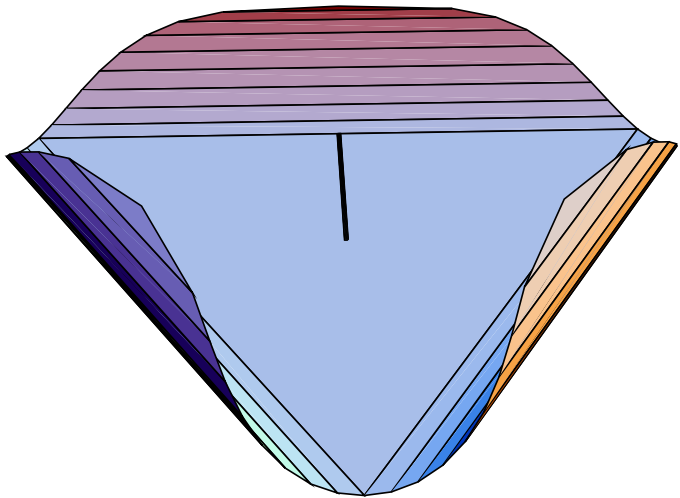}}
  \vspace{0.0cm}
\caption{ A disk with both principal curvatures zero at the
center. The inner triangle is the region where both the principal
curvatures are zero. The three curved regions have one non-zero
principal curvature. The straight line are the generators of this
surface. The thick line is a generator through the center that can not
be extended till the boundary.  The inner triangle is the set ${\cal
A}_1 = {\cal B}_1$ and the three curved regions together are the set
${\cal B}_0$. The generators in the the curved regions are the {\em
leaves} of the relative nullity distribution $\Delta$ in ${\cal B}_0$
and the inner triangle is a leaf of $\Delta$ in ${\cal B}_1$. (See
Sec.~\ref{sec:proofs} for the definitions of ${\cal A}_1, {\cal B}_1,
{\cal B}_0, \Delta$ and ``the leaves of a distribution''.)  }
\label{fig:tooflat}       
\end{figure}

\begin{figure}[tbh]
\epsfxsize = 0.7 \hsize
\centerline{\epsfbox{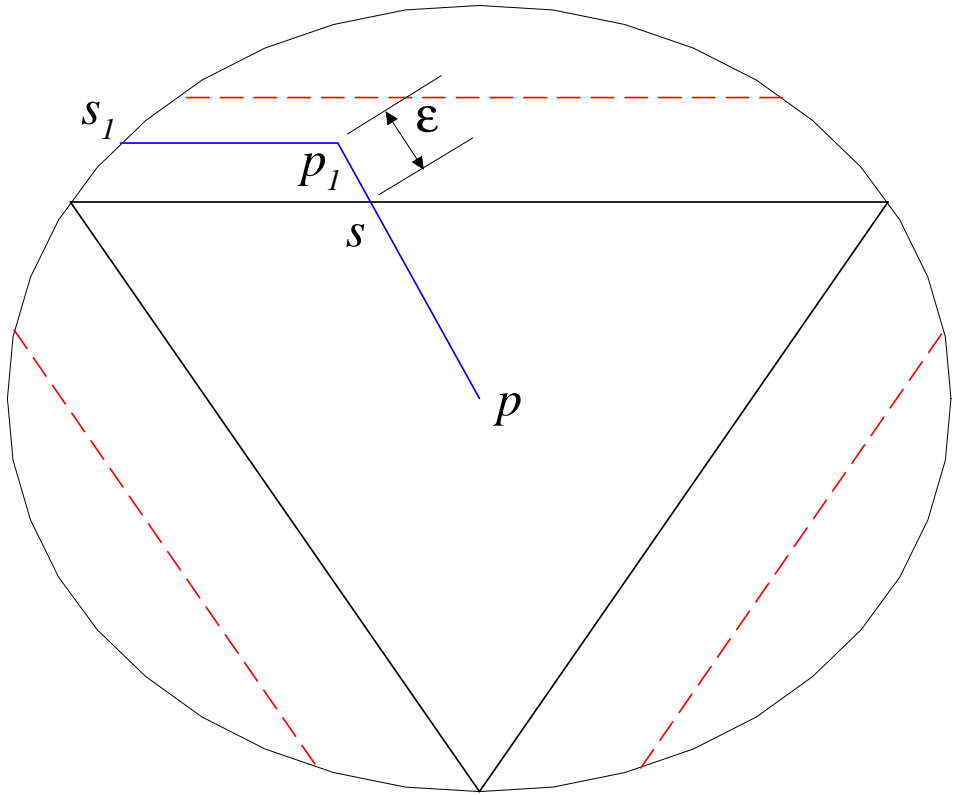}}
\vspace{0.0cm}
\caption{ This figure is a 2-dimensional representation of the
embedding in Fig.~\ref{fig:tooflat}. The inner triangle is the region
where both the principal curvatures are zero and the three sets
bounded by a circular arc and an edge of the triangle are the regions
with one non-zero principal curvature. The dashed lines represent the
generators in these regions.  The point $p$ is the center of the disk
and $ps$ is a straight line through the center that cannot be extended
keeping it straight in the embedding space.  The point $p_1$ is in the
curved region within $\epsilon$ of $s$.  The point $s_1$ is the
intersection of the generator through $p_1$ with the boundary of the
disk.  }
\label{fig:brkn_sgmnt}       
\end{figure}

We now consider the remaining case: that {\it both} principal
curvatures vanish at $p$.  In this case, there isn't a unique
generator through $p$.  Further, not every local generator through $p$
need be extendible as a straight line in the embedding space
($\mathbb{R}^3$) all the way to the boundary as is evident from
Fig.~\ref{fig:tooflat}. In this case, we pick a local generator and
extend it as far as possible while keeping it a straight line in
$\mathbb{R}^3$. Let the maximal straight line be $ps$.  Both principal
curvatures are zer o at the point $s$, but, given an $\epsilon > 0$,
we can find a point $p_1$ close to $s$ such that one principal
curvature is non-zero at $p_1$ and $d(s,p_1) < \epsilon$ (see
Fig.~\ref{fig:brkn_sgmnt}).  Since the point $p_1$ has one nonzero
principal curvature, from the argument in the previous paragraph, it
follows that there exists a point $s_1$ on the boundary such that the
image of the geodesic $p_1s_1$ in the sheet is a straight line in
$\mathbb{R}^3$. If the curvature is everywhere bounded, by making
$d(s,s_n)$ sufficiently small, we can make the angles between the
segments $p_1s_1$ and $ps$ virtually the same in both the sheet and in
the space so that $d(p,s_1) - r(p,s_1)$ is as small as we please.

We will now show this rigorously. The difference between $d(p,s_1)$
and $r(p,s_1)$ is due to two factors. One contributing factor is that
the small segment $sp_1$ is flat in the sheet but curved in the
embedding space, so that $r(s,p_1) \neq d(s,p_1)$. However, both these
lengths are small, since they are bounded by $\epsilon$, and the
contribution of this segment to $d(p,s_1) - r(p,s_1)$ is bounded by $2
\epsilon$.

Another contributing factor is that, since the curvature is not zero
on the segment $sp_1$, the angles between the straight segments $ps$
and $p_1s_1$ in the embedding space and in the sheet are not
equal. Let $\theta$ denote the angle between the segments $ps$ and
$p_1s_1$ in the sheet and let $\theta'$ be the angle between the
images of these segments in the embedding space. Since the curvature
is a measure of the rate of change of the angle that a tangent vector
makes with respect to a fixed axis, it is easily seen that
$$
|\theta - \theta'| \leq K d(s,p_1),
$$ 
where $K$ is a bound on the components of the curvature tensor
${\mathbf K}^a_{\alpha \beta}$.  Since the segments $ps$ and $p_1s_1$
are straight both in the sheet and in the embedding space, $r(p,s) =
d(p,s)$ and $r(p_1,s_1) = d(p_1,s_1)$.  Also, $r(s,p_1) \leq d(s,p_1)
\leq \epsilon$.  Combining the estimates
$$ 
|d(p,s_1) - \sqrt{d(p,s)^2 + d(p_1,s_1)^2 - 2 d(p,s) d(p_1,s_1)
  \cos(\theta)}| \leq d(s,p_1) \leq \epsilon,
$$
$$ 
|r(p,s_1) - \sqrt{r(p,s)^2 + r(p_1,s_1)^2 - 2 r(p,s) r(p_1,s_1)
  \cos(\theta')}| \leq r(s,p_1) \leq \epsilon,
$$
$$
|\cos(\theta) - \cos(\theta')| \leq |\theta - \theta'|
$$
with the results from above and using $d(p,s) \leq R, d(p_1,s_1) \leq
R$, where $R$ is the radius of the disk, we obtain
$$
|d(p,s_1) - r(p,s_1)| \leq C_1 (2+ K R) \epsilon
$$ 
where $C_1 > 1$ is a constant. Note that the bound on the right hand
side reflects the contributions of both the factors -- the term $2
\epsilon$ is the contribution from the length of the segment $sp_1$,
and the term $K R \epsilon$ is the contribution from the difference
between the angles $\theta$ and $\theta'$.

Now, we choose a sequence $\delta_n \rightarrow 0$ and repeat this
construction with $\epsilon = \delta_n$, for each $n$.  This will
yield a point $s_1(n) = q_n$ on the boundary with $r(p,q_n) - d(p,q_n)
\leq C_1 (2+ K R) \delta_n $ so that $r(p,q_n) - d(p,q_n) \rightarrow
0$ as $n \rightarrow \infty$.  Since the boundary of the unit disk is
compact, the sequence $q_n$ has an accumulation point $q$ on the
boundary and there exists a subsequence $q_{n_k}$ that converges to
$q$. From the above estimate, it follows that $r(p,q) = d(p,q)$ and
$d(p,q) = 1$ implies that the disk cannot be isometrically embedded
inside a spherical shell with a diameter less than $1$.

Our generalization to an $m$-disk in a $d$-shell, presented in
Sec.~\ref{sec:thm}, is along the same lines as the argument above for
developable surfaces in $\mathbb{R}^3$. Now the local curvature is a
vector-valued tensor rather than a scalar-valued tensor, so that there
are no obvious ``principal curvatures'' and no obvious generalization
of the Gaussian curvature.  Nevertheless, we will show that there are
analogs of the local generators. In the situation $d < 2m$,
Lemma~\ref{lem:kernel} along with Corollary~\ref{cor:null} asserts
that there exists at least one line through the point $p$ in the
$m$-sheet that has zero curvature in the embedding space at $p$. The
direction of this line is a local ``flat direction'' at $p$.

Lemmas~\ref{lem:open},~\ref{lem:dense} and \ref{lem:complete} show
that there exists a ``large'' set of points ${\cal U}$ such that --
\begin{enumerate}
\item Almost every point in the sheet is in ${\cal U}$.
\item If $p \in {\cal U}$, it is possible to extend the line that has
zero curvature locally to get a straight line of finite length that is
also a geodesic in the embedding space.
\end{enumerate}
Given a point $p_0 \in {\cal U}$, there exists a maximal extension of
a zero curvature line through $p_0$, and we will denote the other
endpoint of this segment by $s_0$. Therefore, $s_0$ has the property
that it is not possible to extend $p_0s_0$ any further while keeping
it a geodesic in the embedding space. Lemma~\ref{lem:complete} also
tells us that the number of flat directions is a constant along a
geodesic so that it is the same at $s_0$ and $p_0$.

Given any $\epsilon > 0$, Lemma~\ref{lem:baby_thm} asserts the
existence of a point $p_1$ in ${\cal U}$ such that $d(s_0,p_1) <
\epsilon$ and the number of local flat directions at $p_1$ is less
than the number of flat directions at $s_0$. Since $p_1 \in {\cal U}$,
we can now start from $p_1$ and proceed along a maximally extended
geodesic until its termination point $s_1$. Repeating this process, we
get sequences of points $p_2,p_3,\ldots,p_k$ and $s_1,s_2,\ldots,s_k$
such that
\begin{enumerate}
\item The image of
the straight line $p_js_j$ is a geodesic in the embedding space.
\item  $p_{j+1}$ has fewer local flat directions than $p_{j}$, and
\item $d(s_j,p_{j+1}) < \epsilon$.
\end{enumerate}  
Since the number of flat directions is between $1$ and $m$ everywhere
on the sheet, it follows that the sequences $p_j$ and $s_j$ are
finite.  Consequently, for some finite $k$, the point $s_k = q$ lies
on the boundary of the disk.

Lemma~\ref{lem:flat} shows that for the image
of a straight line $pq$ in the sheet to be a geodesic in the embedding
space, it is necessary and sufficient that $r(p,q) = d(p,q)$. It
follows that $r(p_j,s_j) = d(p_j,s_j)$ for all
$j$. Lemma~\ref{lem:trans} shows that if $r(p,p_1) = d(p,p_1)$ and
$r(p_1,p_2) = d(p_1,p_2)$ , then $r(p,p_2) = d(p,p_2)$. That is, if
two points $p$ and $q$ can be connected by a curve that is a piecewise
geodesic in the embedding space, the image of the straight line
connecting $p$ and $q$ in the sheet is also a geodesic in the
embedding space.

By taking a sequence $\delta_n \rightarrow 0$, we can repeat the above
construction for each $n$ with $\epsilon = \delta_n$. This will yield
a sequence of points $q_n$ on the boundary such that $p$ can be
connected to $q_n$ by a discontinuous, piecewise straight curve, with
a finite number of straight line segments $p_js_j$. The
discontinuities go to zero as $n$ goes to infinity since
$d(s_j,p_{j+1}) < \delta_n$. The compactness of the boundary of the
disk, gives an accumulation point $q$ for the sequence $q_n$. The
above construction along with Lemma~\ref{lem:flat} and
Lemma~\ref{lem:trans} show that, given any point $p$ in the disk,
there exists a point $q$ on the boundary of the disk such that $r(p,q)
= d(p,q)$. Since $d(p,q) = 1$ when $p$ is the center of the disk, it
follows that the disk cannot be embedded in a ball with diameter less
than 1.

\section{The geometry of a confined $m$-sheet} \label{sec:thm}

We will consider the problem of the existence of an isometry $\phi :
D^m \rightarrow \bar{B}_r$. In Sec.~\ref{sec:rvw}, we will
set up the notation we use and review the differential geometry of
isometric immersions. In Sec.~\ref{sec:proofs}, we prove the results
stated in Sec.~\ref{sec:problem}.

\subsection{Review of Differential Geometry} \label{sec:rvw}

In this section, we will review the differential geometry of the
isometric immersion of $m$-manifolds in $d$-manifolds. We will use the
co-ordinate (index) free language and follow the presentation in
Ref.~\cite{daj}, but we will restate our results in the indexed
notation where appropriate.

We will consider the case of a smooth $m$-manifold $M$, typically the
unit disk in $m$ dimensional Euclidean space, immersed in a smooth
$d$-manifold $\tilde{M}$, typically $R^d$.  Since both $M$ and
$\tilde{M}$ are subsets of Euclidean spaces, we can find global
Cartesian co-ordinates (a co-ordinate patch that covers the entire
manifold) for both $M$ and $\tilde{M}$. We will denote these
co-ordinates on $M$ by $x^{\alpha}$ (greek superscripts) and on
$\tilde{M}$ by $r^a$ (roman superscripts).

We will require that the immersion $\phi : M \rightarrow \tilde{M}$ be
smooth. Here, $\phi$ is the co-ordinate free representation of the
immersion given by the $d$ functions $r^a(x^{\alpha}), a =
1,2,\ldots,d$. Let $TM$ and $T\tilde{M}$ denote the tangent bundles of
$M$ and $\tilde{M}$ respectively.  We will use $(\psi)^a$ with a roman
superscript to denote the indexed representation of a quantity $\psi$
that takes values in $\tilde{M}$ or $T\tilde{M}$ and $(U)^{\alpha}$
with a greek superscript to denote the indexed representations of a
quantity $U$ that take values in $M$ or $TM$.  Using this notation,
the statement that $\phi(x)$ is the index free representation of 
$r^a(x^{\alpha})$ is written as $(\phi)^a((x)^{\alpha}) =
r^a(x^{\alpha})$.

The immersion induces a map $\phi_*(x) : T_x M \rightarrow T_{\phi(x)}
\tilde{M}$ between the tangent space of $M$ at $x$ and the tangent
space of $\tilde{M}$ at $\phi(p)$ that is injective (one to one) for
each $x \in M$.  Since the immersion $\phi$ is isometric,
$$
\langle \phi_*X,\phi_*Y \rangle_{\tilde{M}} = \langle X,Y \rangle_M, 
$$ 
for every $x \in M$ and $X,Y \in T_x M$, $\langle.,.\rangle_M$ denotes
the Riemannian metric on $M$ and similarly for $\tilde{M}$.  For the
case $\phi:D^m \rightarrow \mathbb{R}^d$ that we are considering,
since the $x^{\alpha}$ are global Cartesian co-ordinates, the metric
on $D^m$ is given by $\langle X,Y \rangle_M = \delta_{\alpha
  \beta}(X)^{\alpha} (Y)^{\beta}$, and the immersion is given by a
vector valued function $\mathbf{r}(x^{\alpha})$ with Cartesian
components $r^a(x^{\alpha})$. The above equation is then equivalent to
the statement
$$
\partial_{\alpha} \mathbf{r} \cdot \partial_{\beta} \mathbf{r} =
\delta_{\alpha \beta}.
$$ 
We are therefore asserting that the strain $u_{\alpha \beta}$
is identically zero.

Locally identifying $M$ with its image under $\phi$, we can consider
the tangent space of $M$ at $x$ as a subspace of the tangent space of
$\tilde{M}$. Because of this identification, we will henceforth denote
the vector $\phi_*X \in T\tilde{M}$ by $X$ for all $X \in TM$. For
each point $x \in M$, we will denote the subspace of vectors in
$T_{\phi(x)}\tilde{M}$ that are orthogonal to the vector $X = \phi_*X
\in T\tilde{M}$ for all $X \in TM$ by $T_x M^{\perp}$. This gives the
decomposition
$$
T_{\phi(x)} \tilde{M} = T_x M \oplus T_x M^{\perp},
$$
where $T_x M^{\perp}$ is the orthogonal complement of $T_x M$ in
$T_{\phi(x)} \tilde{M}$. The vector bundle 
$$
TM^{\perp} = \bigcup_{x \in M} T_x M^{\perp},
$$ 
is called the normal bundle to $M$. Therefore, the union of the
tangent spaces $T_y\tilde{M}$ over all the points in $y \in \tilde{M}$
that are in the image $\phi(M)$, is the vector bundle given by
$$ \left. T\tilde{M} \right| _{\phi(M)} = \bigcup_{x \in M} \left(T_x
M \oplus T_x M^{\perp}\right).
$$ 
This will motivate defining the tangential projection
$$
()^T : \left. T\tilde{M} \right| _{\phi(M)} \rightarrow TM,
$$
and the normal projection
$$
()^{\perp} : \left. T\tilde{M} \right| _{\phi(M)} \rightarrow TM^{\perp}.
$$

As explained in the Appendix, there exists a differential operator on
a Riemannian manifold, called the {\em Levi-Civita} connection
\cite{waldbook}.The connections on $\tilde{M}$ and $M$ are related by
the {\em Gauss Formula}:
\begin{equation}
\tilde{\nabla}_X Y = (\tilde{\nabla}_X Y)^T + (\tilde{\nabla}_X
Y)^{\perp} = \nabla_X Y + \alpha(X,Y), \label{eq:gauss1}
\end{equation}
where $X,Y \in TM$ are arbitrary (smooth) vector fields and $\alpha:
TM \times TM \rightarrow TM^{\perp}$ is a symmetric bilinear map.
$\alpha$ is called the {\em second fundamental form} of $\phi$.  A
calculation using the representation $r^a(x^{\alpha})$ for the
immersion $\phi$ shows that
$$ (\alpha(X,Y))^a = K^a_{\alpha \beta} (X)^{\alpha} (Y)^{\beta}.
$$ 
$\alpha$ is therefore the co--ordinate free representation of the extrinsic
curvature ${\mathbf K}_{\alpha \beta}$ (See Eq.~(\ref{eq:extr_curv})).

Given vector fields $X \in TM$ and $\xi \in TM^{\perp}$, the shape
operator $A_{\xi}: TM \rightarrow TM$ is defined by
$$
A_{\xi} X = -(\tilde{\nabla}_X \xi)^T.
$$
Using $\langle \xi, Y \rangle = 0$ for every vector field $Y \in TM$
and computing $\tilde{\nabla}_X \langle \xi, Y \rangle$ yields
$$
\langle A_{\xi} X,Y \rangle = \langle \alpha(X,Y),\xi \rangle.
$$ 
We denote the normal component of $\tilde{\nabla}_X \xi$ by
$\nabla^{\perp}_X \xi$ and this defines a compatible connection on the
normal bundle. The above equation yields the {\em Weingarten Formula}
\begin{equation}
  \label{eq:weingart}
  \tilde{\nabla}_X \xi = -A_{\xi} X + \nabla_X^{\perp} \xi.
\end{equation}

A characterization of the intrinsic geometry of a manifold is given by
the Riemann Curvature \cite{manifolds,waldbook} --

\begin{defn}For a Riemannian manifold $M$ with a Levi-Civita
  connection $\nabla$, the {\em Riemann} (or {\em Intrinsic}) {\em
    curvature} tensor $R$ is given by
$$
R(X,Y)Z = \nabla_X \nabla_Y Z - \nabla_Y \nabla_X Z -
\nabla_{[X,Y]} Z
$$ where $X,Y,Z \in TM$ are arbitrary (smooth) vector fields and the
{\em Lie-Product} $[.,.]$ is defined by
$$
[X,Y] = \nabla_X Y - \nabla_Y X.
$$
\end{defn} \label{defn:curv}

\begin{rem} The Riemann curvature is zero for the Euclidean space
  ${\mathbb R}^m$. It can be shown that the Riemann curvature is also
  zero for all manifolds that are isometric embeddings of ${\mathbb
  R}^m$ in a larger space.  The converse is also true, so that the
  Riemann curvature of an $m$-manifold $M$ being zero implies that
  given $q \in M$ there is an open set $U \in M$ containing $q$ such
  that there exists an isometry $\phi : V \rightarrow U$, where $V$ is
  an open neighborhood of the origin in ${\mathbb R}^m$.
\end{rem}

Computing the curvature tensor of $\tilde{M}$ and taking the
tangential projection of $\tilde{R}(X,Y)Z$ yields the {\em Gauss
Equation}
\begin{equation}
  \label{eq:gauss2}
  \langle R(X,Y)Z,W \rangle = \langle \tilde{R}(X,Y)Z,W \rangle +
  \langle \alpha(X,W),\alpha(Y,Z) \rangle - \langle
  \alpha(X,Z),\alpha(Y,W) \rangle,
\end{equation}
where $R$ and $\tilde{R}$ are the curvature tensors of $M$ and
$\tilde{M}$ respectively. Taking the normal projection gives the {\em
Codazzi Equation}
\begin{equation}
  \label{eq:codazzi}
  (\tilde{R}(X,Y)Z)^{\perp} = (\nabla^{\perp}_X \alpha)(Y,Z) -
  (\nabla^{\perp}_Y \alpha)(X,Z),
\end{equation}
where
$$
(\nabla^{\perp}_X \alpha)(Y,Z) = \nabla^{\perp}_X (\alpha(Y,Z)) -
\alpha(\nabla_X Y,Z) - \alpha(Y,\nabla_X Z).
$$

\begin{defn}
  A subspace valued function, that is a function defined on a manifold
  $M$, whose value at a point $x \in M$ is a subspace of  $T_xM$ is
  called a {\em distribution}.

  Given an isometric immersion $\phi : M \rightarrow \tilde{M}$, the
  second fundamental form $\alpha : TM \times TM \rightarrow
  TM^{\perp}$ determines a distribution as follows. For each $x \in
  M$, a subspace $\Delta(x)$ is defined by
$$
\Delta(x) = \{ p \in T_xM : \alpha(Y,p) = 0 \mbox{ for all } Y \in T_x
M \}.
$$ 

$\Delta(x)$ is called the {\em subspace of relative nullity} and its
dimensionality is the index of relative nullity $\nu(x)$.  The {\em
  relative nullity distribution} $\Delta$ is the function $x \mapsto
\Delta(x)$, the subspace of relative nullity at $x$.
\end{defn}

\begin{defn} A distribution $D$ is {\em smooth} on a set $U \subseteq M$ if
  there exist smooth vector fields $X_i : U \rightarrow TU, i =
  1,2,\ldots,p$ such that at each $x \in U$, the vectors $X_i(x) \in
  TU, i = 1,2,\ldots,p$ form a basis for the subspace $D(x)$.
\end{defn}
 
\subsection{Immersion Theorems} \label{sec:proofs}

We will now restrict our attention to smooth isometric mappings $\phi
: D^m \rightarrow \mathbb{R}^d$, where $D^m = \{\mathbf{x} \in
\mathbb{R}^m : \|\mathbf{x}\| \leq 1\}$ is the closed unit disk in
$\mathbb{R}^m$, and both $D^m$ and $\mathbb{R}^d$ are equipped with
the standard Euclidean metric. In this case, both $D^m$ and ${\mathbb
R}^d$ have zero Riemann curvature, so that, for the isometric
immersion $\phi$, the Gauss equation (\ref{eq:gauss2}) yields
\begin{equation}
  \label{eq:gauss_flat}
\langle \alpha(X,W),\alpha(Y,Z) \rangle - \langle
\alpha(X,Z),\alpha(Y,W) \rangle = 0
\end{equation}
and the Codazzi equation (\ref{eq:codazzi}) yields
\begin{equation}
  \label{eq:codz_flat}
 (\nabla^{\perp}_X \alpha)(Y,Z) - (\nabla^{\perp}_Y \alpha)(X,Z) = 0,  
\end{equation}
for arbitrary $X,Y,Z,W \in TD^m$.

We will define $\rho(\phi)$, the span or the diameter of the immersion
$\phi$ as
$$
\rho(\phi) = \sup_{\mathbf{x},\mathbf{y} \in D^m}r({\mathbf
  x},\mathbf{y}),
$$
where
$$
r({\mathbf
  x},\mathbf{y}) = \| \phi(\mathbf{x}),  \phi(\mathbf{y})
\|_{\mathbb{R}^d},
$$ is the Euclidean distance between the images of the points
$\mathbf{x}$ and $\mathbf{y}$ in $\mathbb{R}^d$. We will denote the
Euclidean distance in $D^m \subset R^m$ between the points ${\mathbf
  x}$ and $\mathbf{y}$ by $d(\mathbf{x},\mathbf{y})$.

Using this definition of the span of an embedding, we can reformulate
the question about the existence of an isometry $\phi : D^m
\rightarrow \bar{B}_r$ for arbitrary $r > 0$ (See
Sec.~\ref{sec:problem}) as follows -- {\it Given any $R_s$ with
  $0 < R_s < 1$, Is there a smooth isometric immersion $\phi : D^m
  \rightarrow \mathbb{R}^d$, with $\rho(\phi) \leq R_s$?}

We first recall the earlier result \cite{Kramer.PRL} on the existence
of such an isometry for all $R_s > 0$ if $d \geq 2m$.

\begin{theorem} If $d \geq 2m$, there exists a smooth isometry $\phi$ with
  $\rho(\phi) \leq R_s$, for any given $R_s$, $0 < R_s < 1$.
\label{thm:crssring}
\end{theorem}

\begin{proof} This proof is from Ref.~\cite{Kramer.PRL}. Define
$$ 
\phi: (x_1, x_2,\ldots, x_m) \rightarrow (y_1, y_2, \ldots,
  y_{2m}, \ldots y_n),
$$
by
$$
y_{2i -1} = \frac{R_s}{\sqrt{2m}} \cos\left(\frac{\sqrt{2m} x_i}{R_s}
\right), y_{2i} = \frac{R_s}{\sqrt{2m}} \sin\left(\frac{\sqrt{2m}
  x_i}{R_s} \right),
$$
for $i = 1,2,\ldots m$, and
$$
 y_k = 0,
$$ 
for $k > 2m$.  Clearly $\rho(\phi) \leq R_s$. Computing the strain
 $u_{\alpha \beta}$ by Eq.~(\ref{eq:strain}) gives $u_{\alpha \beta} =
 0$ showing that $\phi$ is a smooth isometry with $\rho(\phi) \leq
 R_s$.
\end{proof}

The following theorem is the main result of this paper --

\begin{theorem} If $d < 2m$, and $\phi: D^m \rightarrow \mathbb{R}^d$
  is a smooth isometric immersion, $\rho(\phi) \geq 1$.
\label{thm:crumple}
\end{theorem}
This theorem implies the nonexistence of a smooth isometric immersion
$\phi$ with $\rho(\phi) < R_s$ for arbitrarily small but positive
$R_s$. Before we prove this theorem, we will prove a few results that
are useful in the proof of the theorem.

\begin{defn} Let $V$ and $U$ be finite dimensional vector spaces and
  $\beta: V \times V \rightarrow U$ be a bilinear map. We denote by
  $N(\beta)$ the subspace
$$
N(\beta) = \{ n \in V: \beta(Y,n) = 0, \mbox{for all } Y \in V \}
$$ 
called the (right) {\em kernel} of $\beta$. We may define,
similarly the left kernel. If $\beta$ is symmetric, the left and the
right kernels agree and the subspace $N(\beta)$ is called the {\em
  kernel} of $\beta$.
\end{defn}

\begin{defn} Let $V$ and $U$ be finite dimensional vector spaces. A 
  bilinear map $\beta : V \times V \rightarrow U$ is {\em flat} with
  respect to a non-degenerate inner product $\langle.,.\rangle : U
  \times U \rightarrow \mathbb{R}$ if
$$
\langle \beta(X,W),\beta(Y,Z) \rangle - \langle
\beta(X,Z),\beta(Y,W) \rangle = 0
$$ 
for all $X,Y,Z,W \in V$.
\end{defn}

\begin{lemma} Let $\beta : V \times V \rightarrow U$ be a symmetric
  flat bilinear form with respect to the positive definite inner
  product $\langle.,.\rangle : U
  \times U \rightarrow \mathbb{R}$. Then
$$
\dim N(\beta) \geq \dim V - \dim U.
$$
\label{lem:kernel}
\end{lemma} 

This result and the corollary that follows (Corollary~\ref{cor:null})
are due to E. Cartan~\cite{cartan}.

\begin{proof}
  
Setting $W = X$ and $Z = Y$ in the definition of a flat bilinear
form and using the symmetry of $\beta$ gives
$$ 
  \langle \beta(X,X),\beta (Y,Y) \rangle = \langle \beta(X,Y),
  \beta(X,Y) \rangle 
  \hspace{5pc} (*)
$$ 
for every $X,Y \in V$. We will use this result repeatedly in the proof
of the lemma.

Let $\dim V = m$ and $\dim U = k$. We need to demonstrate the
existence of at least $m - k$ linearly independent vectors $n_1, n_2,
\ldots, n_{m-k}$ such that $\beta(n_i,Y) = 0$ for all $Y \in V$ and $i
= 1,2,\ldots,m-k$.

We only need to consider the case $m > k$ as there is nothing to prove
for $m \leq k$. It suffices to demonstrate the existence of a single
$p \neq 0$ satisfying $\beta(Y,p) = 0$ for all $Y \in V$.  To see
this, note that, having found one such $p$, $\beta$ naturally induces
a map $\beta_1 : V/V_1 \times V/V_1 \rightarrow U$ where $V_1$ is the
subspace spanned by $p$ so that $V/V_1$ is a vector space of dimension
$m-1$ and then repeat the demonstration of the existence of a vector
$p_1 \in V/V_1$.  Continue in this way, reducing the dimension of the
quotient by one at each step, until that dimension has been reduced to
$k$.  In fact, it suffices to find a nonzero $p$ satisfying $
\beta(p,p)= 0$, for this automatically implies $\beta(Y,p)= 0$ for all
$Y \in V$ by $(*)$ and the positive-definiteness of $\langle Z,
Z\rangle$.

To exhibit a suitable $p$, we first construct a maximal set of
non-zero vectors $p_1,p_2,\ldots,p_l$ such that $\beta(p_i,p_j) = 0$
for $i \neq j$. We will construct this set inductively.  Let
$p_1,\cdots,p_s$ be a set of $s$ elements of $V$, such that
$\beta(p_i,p_j)=0$ for $i,j = 1,\cdots, s;\ i\neq j$. Such a set
always exists since we can choose $s = 1$ and set $p_1$ to be any
non-zero vector in $V$. This set can be enlarged since we can show the
existence of an $(s+1)$th vector, $p_{s+1}$, such that the above holds
for $i,j = 1,\cdots, (s+1);\ i\neq j$.  To this end, set $v =p_1
+\cdots +p_s$.  Choose any $q \neq 0$ in $V$ with $\beta(q,v)= 0$.
This is always possible, since $\dim V > \dim U$, and to find such a
$q$, we need to solve $k$ linear equations in $m > k$ unknowns.  Then
$(*)$, with $ X = q$, $Y = v$, yields
\begin{displaymath}
0 = \langle \beta(q,q) , \beta(v,v) \rangle 
= \langle \beta(q,q) , (\sum \beta(p_i,p_i)), \rangle 
\end{displaymath}
where, we have used the bilinearity of $\beta(.,.)$ and
$\langle.,.\rangle$. But for each $i$, we have
\begin{displaymath}
\langle \beta(q,q) , \beta(p_i,p_i) \rangle = \|\beta(q,p_i) \|^2 
\geq 0,
\end{displaymath}
by $(*)$ and the positive-definiteness of $\langle.,.\rangle$.  It
follows that $\beta(q,p_i) = 0$ for all $i=1,\cdots,s$.  So,
$p_{s+1}=q$ is the desired vector.

Using $(m+1)$ times the result of the paragraph above, construct
nonzero vectors $p_1,\cdots,p_{m+1}$, with $\beta(p_i,p_j)= 0$, for
$i,j=1,\cdots,(m+1);\ i\neq j$.  These must be dependent, i.e., we
must have $\sum r_ip_i = 0$, with at least one $r_i$, say $r_n$,
nonzero.  But now we have $0=\beta(p_n,\sum r_i p_i) =
r_n\beta(p_n,p_n)$. We conclude that $ \beta(p_n,p_n)=0$. Setting $p =
p_n$ gives the existence of a $p$ such that $\beta(p,p) = 0$ and as
remarked above, this suffices to prove the lemma.

\end{proof}

Let $\phi: D^m \rightarrow \mathbb{R}^d$ be a smooth isometric
immersion. The second fundamental form $\alpha$ of the immersion
$\phi$ at is a symmetric bilinear map $\alpha: TM \times TM
\rightarrow TM^{\perp}$. Eq.~(\ref{eq:gauss_flat}) implies that
$\alpha$ is a flat bilinear form. At each point $\mathbf{x} \in D^m$,
where $TM_\mathbf{x}$ is a vector space with dimension $m$ and
$TM^{\perp}_{\mathbf{x}}$ is a vector space with dimension $d-m$. Also,
the Kernel $N(\alpha_\mathbf{x})$, is the subspace of relative
nullity $\Delta(\mathbf{x})$. Therefore, Lemma~\ref{lem:kernel} yields
the following corollary:

\begin{cor}Let $\phi: D^m \rightarrow \mathbb{R}^d$ be a smooth, isometric
  immersion. Then, for all $\mathbf{x} \in D^m$, the index of
  relative nullity $\nu(\mathbf{x}) \geq 2m - d$.
\label{cor:null}
\end{cor} 

We denote by $\nu_0$ the {\em index of minimum relative nullity} of
$\phi$ given by
$$
\nu_0 = \min_{\mathbf{x} \in D^m} \nu(\mathbf{x}).
$$ 
By Corollary~\ref{cor:null}, we have that $\nu_0 \geq 2m - d$. We
are considering the case $d < 2m$ so that $\nu_0 \geq 1$.

\begin{defn} The indices of relative nullity at various points in
  $D^m$ can be arranged in an increasing sequence $\nu_0 < \nu_1 <
  \ldots < \nu_k$. Clearly, this is a finite sequence since $\nu_k
  \leq m$. We define the sets $D^m = {\cal A}_0 \supseteq {\cal A}_1
  \supseteq \ldots \supseteq {\cal A}_k$, by
$$
{\cal A}_j = \{ \mathbf{x} \in D^m : \nu(\mathbf{x}) \geq \nu_j \}
$$
for $j = 0, 1, \ldots, k$. We also define the sets ${\cal B}_0, {\cal
  B}_1, \ldots , {\cal B}_k$ by 
$$
{\cal B}_j =  \{ \mathbf{x} \in D^m : \nu(\mathbf{x}) = \nu_j \} =
{\cal A}_j - {\cal A}_{j+1}
$$ 
for $j = 0, 1, \ldots, k$ with the convention ${\cal A}_{k+1}$ is
the empty set $\emptyset$.
\end{defn}

Recall that a distribution is a subspace valued function on $M$. We
will now prove some results about the relative nullity distribution
$\Delta$ that is given by ${\mathbf x} \mapsto \Delta({\mathbf x})$.

\begin{lemma} For an isometric immersion $\phi : M \rightarrow
  \tilde{M}$, we have :
\begin{enumerate}
\item The relative nullity distribution $ \Delta$ is smooth on any
  open set $U$ where $\nu$ is a constant,
\item The set ${\cal B}_j$ is open in ${\cal A}_j$, so that, if ${\mathbf
    x} \in {\cal B}_j$, there is an open neighborhood ${\cal N}$ of
  ${\mathbf x}$ such that $\nu$ is a constant on ${\cal N} \cap {\cal
    A}_j$.
\end{enumerate}
\label{lem:open}
\end{lemma}

A closely related result is proved in Dajczer~\cite{daj} (Prop. 5.2).
The proof presented below is essentially the same as the proof in
\cite{daj}.

\begin{proof}

\begin{enumerate}
\item Set $m = \dim M$. Let $\dim \Delta(\mathbf{x}) = \nu
  (\mathbf{x}) = p$ for all points in the open subset $U$.  Let
  $\Delta^\perp(\mathbf{x})$ be the orthogonal complement of
  $\Delta(\mathbf{x})$ in $T_{\mathbf{x}}M$.  Evidently the dimension
  of $\Delta^\perp$ is $m-p$.  On the other hand, we may express
  $\Delta^\perp$ via
$$ \Delta^{\perp}(\mathbf{x}) = \mathrm{span } \left\{ A_{\xi} X :
    \mbox{ for all } X \in T_{\mathbf{x}}M, \xi \in
    T^{\perp}_{\mathbf{x}} M \right\},
$$ 
given $\mathbf{x}_0 \in U$, there exist $X_1, X_2, \ldots, X_{m-p}
\in TM_{\mathbf{x}_0}$ and $\xi_1, \ldots, \xi_{m-p} \in
TM^{\perp}_{\mathbf{x}_0}$ ,such that
$$
\Delta^{\perp}(\mathbf{x}) = \mathrm{span } \left\{ A_{\xi_j} X_j \right\},
$$ 
Take smooth local extensions $X_1, \ldots, X_{m-p} \in TM$ and $\xi_1,
\ldots, \xi_{m-p} \in TM^{\perp}$. Clearly,
$$
\mathrm{span } \left\{ A_{\xi_j} X_j \right\} \subseteq \Delta^{\perp}.
$$
By continuity, the vector fields $A_{\xi_j} X_j$ are linearly
independent in a neighborhood of $\mathbf{x}_0$. $U$ is open so that we can
choose this neighborhood in such a way that $\dim \Delta^{\perp} = m -
\nu(\mathbf{x}) = m-p$. Therefore, in this neighborhood 
$$
\Delta^{\perp} = \mathrm{span } \left\{ A_{\xi_j} X_j \right\}.
$$
This implies that $\Delta^{\perp}$ is a smooth distribution on $U$
and consequently, so is $\Delta$.
\item follows immediately from the above argument by noting that the
  set ${\cal A}_j$ is the set of all $\mathbf{x}$ with
  $\nu(\mathbf{x}) \geq \nu_j$.
\end{enumerate}

\end{proof}

\begin{cor} The sets ${\cal A}_j$ for $j = 1, \ldots, k$ are all closed
  in $D^m$ by Lemma~\ref{lem:open}.
\label{cor:closed}
\end{cor}

\begin{rem}
  To clarify the relationships between these closed and open sets, we
  consider the 3-cornered shape of Fig.~\ref{fig:tooflat}. In this
  example, $k = 1$, $\nu_0 = 1$ and $\nu_1 = \nu_k = 2$.  The set
  ${\cal A}_0$ is the whole disk, and ${\cal A}_1$ is the flat triangular
  region in the middle.  The set ${\cal B}_0$ consists of the three
  curved side regions, while ${\cal B}_1$ is the same as ${\cal A}_1$.
  Evidently ${\cal B}_1$ is open in ${\cal A}_1$, since the two sets are
  the same.  Any neighborhood of a point in ${\cal B}_1$ that fails to
  be be in ${\cal B}_1$ clearly fails to be in ${\cal A}_1$ as well.  On
  the other hand, ${\cal A}_1$ is not open in the whole disk and this
  implies that ${\cal B}_1$ is not open in the disk either.  

  Thus {\em e.g.}, following any straight line from the flat region
  into the curved region, there is an identifiable last flat point
  with $\nu = 2$.  Turning to ${\cal B}_0$, it is open in ${\cal A}_0$,
  which is the whole disk.  This means that if we traverse our line
  from the curved region into the flat region, there is no last point
  with $\nu = 1$.  These statements give the essential content of
  Lemma~\ref{lem:open} and Corollary~\ref{cor:closed} for this example.
\end{rem}

As we saw above, though ${\cal B}_j$ is open in ${\cal A}_j$, it is
not necessarily open in $\mathbb{R}^m$. As evident from
Lemma~\ref{lem:open}, it will be useful to look at open sets in
$\mathbb{R}^m$ that have a constant index of relative nullity. For
later purposes, it will also be important that the union of these open
sets be "large", in the sense that given any point $p$ in $D^m$, we
can find a sequence of points from these sets that converges to
$p$. The following lemma gives the existence of such sets --

\begin{lemma} $\phi : D^m \rightarrow \mathbb{R}^d$ is an isometric
immersion, and $d < 2m$. Then, we have an open (in $\mathbb{R}^m$) set
${\cal U}$ that is dense in $D^m$ such that
$$
{\cal U} = \bigcup_{j=0}^k {\cal O}_j
$$ 
where ${\cal O}_j$ for $j = 0,1,2,\ldots,k$ is possibly empty, is open
in $\mathbb{R}^m$, and is a subset of ${\cal B}_j$.
\label{lem:dense}
\end{lemma}

\begin{proof} Given a set $G \in \mathbb{R}^m$, let $\partial G$
denote the boundary of $G$, $\bar{G}$ denote the closure of $G$ and
$G^0$ denote the interior of $G$.

First, we note that $A \subseteq D^m$ is closed in $\mathbb{R}^m$
implies that $\partial A$ is nowhere dense. This can be shown as
follows:

Since $A$ is closed, $\partial A \subseteq A$. Therefore, $(\partial
A)^0 \subseteq A^0$. However, $\partial A = \bar{A} - A^0 = A -
A^0$. Consequently, it follows that $(\partial A)^0 = \emptyset$.

Define ${\cal U}$ as 
$$ 
{\cal U} = (D^m)^0 - \bigcup_{j=0}^k
\partial {\cal A}_j = D^m - \partial D^m - \bigcup_{j=0}^k \partial
{\cal A}_j 
$$
Clearly, ${\cal U}$ is open. $D^m$ is a closed subset of
$\mathbb{R}^m$ with a nonempty interior, so that it is a second
category set. Corollary~\ref{cor:closed} implies that each ${\cal
A}_j$ is closed in $\mathbb{R}^m$. Therefore, the preceding argument
along with the Baire category theorem implies that ${\cal U}$ is dense
in $D^m$.

Now, we set ${\cal O}_j = {\cal U} \cap {\cal B}_j$. Clearly, ${\cal
O}_j \subseteq {\cal B}_j$. Also,
$$
\bigcup_{j=0}^k {\cal B}_j = D^m,
$$
implies that 
$$
\bigcup_{j=0}^k {\cal O}_j = {\cal U}.
$$ 

Assume that ${\cal O}_p$ is nonempty and let $q \in {\cal O}_p$ be an
arbitrary point. Then, $q \in {\cal B}_p$. Therefore, $q \in {\cal
A}_p - {\cal A}_{p+1} = {\cal A}_p \cap {\cal A}_{p+1}^C$, where
${\cal A}_{p+1}^C$ is the complement of ${\cal A}_{p+1}$ and is open
in $\mathbb{R}^m$. From the definition of ${\cal U}$, it is clear that
$q \notin \partial {\cal A}_p$. Since ${\cal A}_p$ is closed,
${\cal A}_p^0 = {\cal A}_p - \partial {\cal A}_p$. Therefore, we
have
$$ 
q \in {\cal A}_p^0 \cap {\cal A}_{p+1}^C \cap {\cal U}
\subseteq {\cal B}_p \cap {\cal U} = {\cal O}_p.  
$$

This implies that $q$ is an interior point since ${\cal A}_p^0\, \cap
\,{\cal A}_{p+1}^C \,\cap \,{\cal U}$ is an intersection of open sets,
which gives an open neighborhood of $q$ contained in ${\cal
O}_p$. Since $q$ was an arbitrary point in ${\cal O}_p$, it follows
that every point in ${\cal O}_p$ is an interior point so that ${\cal
O}_p$ is open.  \end{proof}

\begin{obs} Note that, in the above proof, we are not guaranteed the
existence a point in ${\cal O}_j$ for a given $j$, so that ${\cal
O}_j$ could be empty for some $j$. It can not be empty for all $j$
because ${\cal U}$, which is the union of the sets ${\cal O}_j$ over
$j = 1,2,\ldots,k$ is dense in $D^m$.

By definition, ${\cal A}_0 = D^m$ and ${\cal B}_0 \neq \emptyset$.  By
Lemma~\ref{lem:open}, ${\cal B}_0 = B \cap D^m$ for some $B$ open in
$\mathbb{R}^m$. Let $q \in B \cap D^m$. Since $B$ is open, it follows
that there is an open neighborhood ${\cal N}$ of $q$ in $B$. It is
also clear that, for any $q \in D^m$, ${\cal N}$ is an open
neighborhood of $q$ implies that ${\cal N} \cap (D^m)^0 \neq
\emptyset$. Clearly, ${\cal N} \cap (D^m)^0 \subseteq {\cal O}_0$,
which is consequently non empty.
\label{obs:non_triv} 
\end{obs}

Recall that a distribution $D$ of index $p$ defined on an open subset
$U \subset M$ is smooth if we can find a basis $X_1({\mathbf x}),
X_2({\mathbf x}), \ldots, X_p({\mathbf x})$ for $R({\mathbf x})$ at
each point ${\mathbf x} \in U$ such that the basis vectors
$X_1,X_2,\ldots,X_p$ vary smoothly with ${\mathbf x}$.

\begin{defn}
  A smooth distribution $D$ defined on a set $U \subset {\mathbb R}^m$
  with index $p$ is said to possess {\em integral submanifolds} (be
  {\em completely integrable}) if there exists a $m - p$ parameter
  family of $p$ dimensional embedded submanifolds of $U$ such that
  every point ${\mathbf x} \in U$ is contained in exactly one of these
  submanifolds and the tangent space of this submanifold coincides
  with $R({\mathbf x})$.

  The integral submanifolds for a distribution $D$ are called the
  {\em leaves} of $D$.
\end{defn} \label{defn:leaves}

\begin{rem} For the embedding represented in Fig.~\ref{fig:tooflat},
  the generators in ${\cal B}_0$ (the regions with one non-zero
  curvature) restricted to the interior of the disk are leaves of the
  relative nullity distribution $\Delta$.  Similarly, the interior of
  the entire inner triangle is a leaf of $\Delta$ in ${\cal O}_1$.
\end{rem}

\begin{defn}
  Let $U \subseteq M$ be an open set and let $X_1,X_2,\ldots,X_p$ be a
  set of smooth vector fields on $U$. The distribution $D$ given by
$$
{\mathbf x} \mapsto \mathrm{Span}(X_1({\mathbf x}),X_2({\mathbf
    x}),\ldots,X_p({\mathbf x}))
$$ 
is said to be {\em involutive}, if at every ${\mathbf x} \in U$, the
Lie-Product $[X_i,X_j]({\mathbf x}) \in R({\mathbf x})$ for all $1
\leq i,j \leq p$.
\end{defn} \label{defn:involute}

A necessary and sufficient condition for a smooth distribution to be
completely integrable is given by Frobenius' theorem~\cite{waldbook}
which states that a smooth distribution is integrable if and only if
it is involutive.

\begin{defn} Let $M$ be a Riemannian manifold. A {\em geodesic} $\gamma :
  [a,b] \rightarrow M$ is a differentiable curve that is a stationary
  point for the functional
  $$ L[\gamma] = \int_a^b \sqrt{\langle \gamma'(t),\gamma'(t)
    \rangle_{\gamma(t)} }\,dt
  $$ where the end points $\gamma(a)$ and $\gamma(b)$ are fixed and
  $\langle.,,\rangle_{\mathbf x} : T_{\mathbf x}M \times T_{\mathbf
    x}M \rightarrow {\mathbb R}$ is the metric on $M$.
\end{defn}

\begin{defn} Let $\tilde{N} \subset N$ be an embedded submanifold of
  the Riemannian manifold $N$. The metric on $N$ then induces a metric
  on $\tilde{N}$. $\tilde{N}$ is {\em totally geodesic} in $N$ if every
  geodesic on $\tilde{N}$ in the induced metric is also a geodesic in
  $N$.
\end{defn}

Let $M$ be a Riemannian manifold with a smooth distribution $D$
defined on an open subset $U \subset M$. To each $X \in D$, we
can associate a map $C_X : D^{\perp} \rightarrow D^{\perp}$
defined by
$$
C_X Y = -P(\nabla_Y X),
$$ where $P: TU \rightarrow D^{\perp}$ is the orthogonal projection.

\begin{lemma}
  Let $\phi : D^m \rightarrow \mathbb{R}^d$ be an isometric immersion,
  and let ${\cal O}_j \subset D^m$ be an open set where the index of
  relative nullity $\nu$ is equal to some constant $\nu_j$. Then, on
  ${\cal O}_j$, we have:
  \begin{enumerate}
  \item The relative nullity distribution $\Delta$ is smooth and
    integrable, and the leaves are totally geodesic in $D^m$ and
    $\mathbb{R}^d$,
  \item If $\gamma:[0,b] \rightarrow D^m$ is a geodesic such that
    $\gamma([0,b))$ is contained in a leaf of $\Delta$, then
    $\nu(\gamma(b)) = \nu_j$.
  \end{enumerate}
\label{lem:complete}
\end{lemma}

This result has been proved by several authors (See {\em e.g.}
\cite{ferus}). We present an outline of the proof in Dajczer
\cite{daj} below. The interested reader will find all the details in
Ref.~\cite{daj} (Thm. 5.3).

\begin{proof}
  \begin{enumerate}
  \item Let $X,Y \in \Delta$ and $Z \in TM$. Then $\alpha(X,Y) = 0$ so
    that $\nabla^{\perp}_Z (\alpha(X,Y)) = 0$, and $\alpha(X,W) =
    \alpha(Y,W) = 0$ for any arbitrary vector $W \in TM$. Therefore,
$$
(\nabla^{\perp}_Z \alpha)(X,Y) = \nabla^{\perp}_Z (\alpha(X,Y)) -
\alpha(\nabla_Z X,Y) - \alpha(X,\nabla_Z Y) = 0.
$$
Using Eq.~(\ref{eq:codz_flat}), we obtain
$$
(\nabla^{\perp}_X \alpha)(Z,Y) =  - \alpha(Z,\nabla_X Y) = 0.
$$ 
This implies that $\nabla_X Y \in \Delta$. Also,
$$
\tilde{\nabla}_X Y = \nabla_X Y + \alpha(X,Y) = \nabla_X Y \in \Delta.
$$ A similar argument shows that $\tilde{\nabla}_Y X = \nabla_Y X \in
\Delta$.  Therefore, $[X,Y] = \nabla_X Y - \nabla_Y X \in \Delta$.  By
the Frobenius' Theorem cited above, $\Delta$ is an integrable
distribution in $D^m$. By a similar argument, $\Delta$ is also
integrable in $\mathbb{R}^d$. Since $\Delta$ is the subspace of
relative nullity, it follows that $\Delta$ is totally geodesic in both
$D^m$ and ${\mathbb R}^d$.
\item 
It can be shown that, given any $W \in T_{\gamma(0)}D^m$, there
  exists a unique vector field $Y$ defined on $\gamma([0,b))$ defined
  by
$$
Y(0) = W; \frac{d}{dt} Y + C_{\gamma'} Y = 0 \mbox{ for } t \in [0,b),
$$ and $Y$ extends smoothly to $t = b$. Let $Z$ be a vector that is
parallel transported along $\gamma$. A calculation shows that
$\|\alpha(Y,Z)\|$ is a constant along $\gamma$. Therefore, if $Z \in
\Delta(\gamma(b))$, $\alpha(Z,Y) = 0$ for all $Y \in
T_{\gamma(b)}D^m$. Therefore, if $W$ is any vector in
$T_{\gamma(0)}D^m$, it follows that $\alpha(Z,W) = 0$, so that $Z \in
\Delta(\gamma(0))$. Consequently, $\nu(\gamma(b)) \leq \nu_j$.  Since
$A_j$ is closed by Corollary~\ref{cor:closed}, $\nu(\gamma(b)) \geq
\nu_j$. Combining these results, we have $\nu(\gamma(b)) = \nu_j$.
\end{enumerate}
\end{proof}

\begin{defn}
  Let $M$ be a Riemannian manifold with a metric given by $\langle ,
  \rangle : TM \times TM \rightarrow \mathbb{R}$. A {\em unit speed}
  curve $\beta: [0,b) \rightarrow M$ is a differentiable curve
  satisfying
$$
\langle \beta'(t), \beta'(t) \rangle_{\beta(t)} = 1,
$$
where
$$ 
\beta'(t_0) = \left. \frac{d}{dt} \beta(t) \right|_{t = t_0} \in
TM_{\beta(t_0)}.
$$
\end{defn}

By appropriately reparameterizing any differentiable curve $\delta :
[0,d) \rightarrow M$, we can obtain a unit speed curve $\beta : [0,b)
\rightarrow M$, with $\beta(t) = \gamma(s(t))$, where $s(t)$ is a
differentiable reparameterization satisfying $s(0) = 0$. Henceforth,
without loss of generality, we will assume that all the curves we
consider, and in particular the geodesics are unit speed.

\begin{obs}
  For every $\mathbf{x} \in {\cal U}$, where ${\cal U}$ is as defined
  in Lemma~\ref{lem:dense}, Lemma~\ref{lem:complete} guarantees the
  existence of at-least one unit speed geodesic $\gamma :[0,b)
  \rightarrow D^m$ with $\gamma(0) = \mathbf{x}$ and $b > 0$ , that is
  contained in a leaf of $\Delta$. We will denote this geodesic by
  $(\gamma, b)$. We can partially order the set of all such geodesics
  by $(\gamma_1,b_1) \leq (\gamma_2,b_2)$ if an only if $b_1 \leq b_2$
  and $\gamma_1(t) = \gamma_2(t)$ for all $t \in [0,b_1)$. It is
  straightforward to verify the existence of at-least one maximal
  geodesic $\gamma^* :[0,b^*) \rightarrow D^m$. Since $\gamma^*$ is
  unit speed, it follows that the limit $ \lim_{b \rightarrow b^*}
  \gamma^*(b) = \mathbf{z}^*$ exists. Also, ${\mathbf z}^* \in D^m$
  since $D^m$ is closed.  

By the maximality of the geodesic $(\gamma^*,b^*)$, either
\begin{enumerate}
\item $\mathbf{z}^* \in (D^m)^0$ and no extension of
  this geodesic is contained in a leaf of $\Delta$, or
\item  $\mathbf{z}^* \in \partial D^m.$
\end{enumerate}
\label{obs:maximal}
\end{obs}

\begin{obs}
Let ${\mathbf x} \in {\cal U}$ and let $\gamma^* :[0,b^*) \rightarrow
D^m$ be a maximal geodesic with $\gamma^*(0) = {\mathbf x}$ as in
Observation~\ref{obs:maximal}. Let $\mathbf{z}^* = \lim_{b \rightarrow
b^*} \gamma^*(b)$. Since every leaf of $\Delta$ is totally geodesic in
both $D^m$ and ${\mathbb R}^d$ and $\gamma^* :[0,b^*) \rightarrow D^m$
is in a leaf of $\Delta$, it follows that $r({\mathbf x},\gamma^*(b))
= d({\mathbf x},\gamma^*(b))$ for all $b \in [0,b^*)$. Taking limits,
we obtain $r({\mathbf x},{\mathbf z}^*) = d({\mathbf x},{\mathbf
z}^*)$.
\label{obs:straight}
\end{obs}

\begin{lemma}
  Given $\mathbf{x} \in {\cal U}$, let $\gamma^* : [0,b^*) \rightarrow
  D^m$ be a maximal geodesic contained in a leaf of $\Delta$ and let
$$
\lim_{b \rightarrow b^*} \gamma^*(b) = \mathbf{z}^* \in D^m.
$$
Then, we have that,
\begin{enumerate}
\item If $\mathbf{z}^* \in (D^m)^0$, given $\epsilon >
  0$, there exists a $\mathbf{w} \in \cal U$ such that
  $d(\mathbf{z}^*,\mathbf{w}) < \epsilon$ and $\nu(\mathbf{w}) <
  \nu(\mathbf{x})$.
\item If $\mathbf{x} \in {\cal O}_0$, $\mathbf{z}^* \in \partial D^m$.
\end{enumerate}
\label{lem:baby_thm}
\end{lemma}

\begin{proof}
\begin{enumerate}
\item Let $\nu(\mathbf{x}) = \nu_k$. By Lemma~\ref{lem:complete},
  $\nu(\mathbf{z}^*) = \nu_{k}$. We see that $\mathbf{z}^* \in {\cal
A}_{k + 1}^C$ which is open by Corollary~\ref{cor:closed}, where, as
before, ${\cal A}_{k+1}^C$ denotes the complement of ${\cal A}_{k+1}$.
Let $C = B(\mathbf{z}^*, \epsilon/2) \cap {\cal A}_{k+1}^C$, where $
B(\mathbf{z}, \epsilon/2) = \{\mathbf{y} \in D^m :
  d(\mathbf{y},\mathbf{z}) < \epsilon/2 \}$ is the open ball with
  radius $\epsilon/2$. Then $C$ is a nonempty open set with
  $\nu(\mathbf{y}) \leq \nu_k$ for all $\mathbf{y} \in C$ . If
  $\nu(\mathbf{y}) = \nu_k$ for all $\mathbf{y} \in C$, by
  Lemma~\ref{lem:complete} $\Delta$ is smooth and integrable on $C$ so
  that we can extend the geodesic beyond $\mathbf{z}^*$ remaining in a
  leaf of $\Delta$, thereby contradicting the maximality of the
  geodesic $(\gamma^*,b^*)$. Therefore, there exists a $\mathbf{y}^*
  \in C$ such that $\mathbf{y}^* \in {\cal A}_k^C$ which is open
  (Corollary~\ref{cor:closed}).  Therefore, $D =
  B(\mathbf{y}^*,\epsilon/2) \cap {\cal A}_k^C$ is a nonempty open
  set.  Now, ${\cal U}$ is dense, so that ${\cal U} \cap D$ is
  nonempty.  Choose any $\mathbf{w} \in {\cal U} \cap D$. By the
  triangle inequality for the metric on $D^m$, we have
$$ d(\mathbf{z}^*,\mathbf{w}) < d(\mathbf{z}^*,\mathbf{y}^*) +
  d(\mathbf{y}^*,\mathbf{w}) < \epsilon/2 + \epsilon/2 = \epsilon.
$$
$\mathbf{w} \in {\cal A}_k^C$ implies that $\nu(\mathbf{w}) <
\nu(\mathbf{x}) = \nu_k$.
\item This follows immediately from the preceding result by
  Observation~\ref{obs:maximal} and the definition of $\nu_0$. 
\end{enumerate}
\end{proof}

\begin{lemma}
  Let $\mathbf{x}, \mathbf{y} \in D^m$ with $d(\mathbf{x},\mathbf{y})
  = r_0$ and let $\gamma:[0,r_0] \rightarrow D^m$ be the unit speed
  geodesic between ${\mathbf x}$ and ${\mathbf y}$. Then,
$$ 
\gamma(t) = \frac{r_0-t}{r_0} \mathbf{x} + \frac{t}{r_0} \mathbf{y}.
$$
$r(\mathbf{x},\mathbf{y}) = r_0$ if and only if the unit speed
curve $\beta : [0,r_0] \rightarrow {\mathbb R}^d$ given by 
$$ 
\beta(t) = \frac{r_0-t}{r_0} \phi(\mathbf{x}) + \frac{t}{r_0}
  \phi(\mathbf{y}),
$$
is such that $\beta = \phi \circ \gamma$.
\label{lem:flat}
\end{lemma}

\begin{proof}
We begin with the following observations which are easily verified
using the fact that geodesics are extremal curves for the arc length.
\begin{enumerate}
\item If $\mathbf{u}, \mathbf{v} \in \mathbb{R}^k$, the unit speed geodesic
  through $\mathbf{u}$ and $\mathbf{v}$ is unique (except for
  reparameterization)  and is given by 
$$ \delta(t) =
  \frac{\|\mathbf{v}-\mathbf{u}\|-t}{\|\mathbf{v}-\mathbf{u}\|}
  \mathbf{u} + \frac{t}{\|\mathbf{v}-\mathbf{u}\|} \mathbf{v}.
$$
\item Every unit speed curve $\eta : [0,b] \rightarrow \mathbb{R}^k$
with $\eta(0) = \mathbf{u}$ and $\eta(b) = \mathbf{v}$ has $b \geq
\|\mathbf{v}-\mathbf{u}\|$ with equality if and only if the curve
$\eta$ is identical to the curve $\delta$ above.
\end{enumerate}

Assume that $r(\mathbf{x},\mathbf{y}) = r_0$. In this case we have
$\eta = \phi \circ \gamma$ is a unit speed curve with $\eta(0) =
\phi(\mathbf{x})$ and $\eta(r_0) = \phi(\mathbf{y})$ so that by item
(ii) above, we have that $\eta$ is a unit speed geodesic in
$\mathbb{R}^d$. By the uniqueness of the geodesic, it follows that
$\beta = \eta$, so that $\beta = \phi \circ \gamma$.

We will now show the converse. Using the standard identification
between ${\mathbb R}^k$ and $T_{\mathbf u} {\mathbb R}^k$, where
${\mathbf u} \in {\mathbb R}^k$ is an arbitrary point, we have
$$
\gamma'(0) = \frac{{\mathbf x} - {\mathbf y}}{r_0},
$$
and 
$$
\beta'(0) = \frac{\phi({\mathbf x}) - \phi({\mathbf y})}{r_0}.
$$
Since $\phi$ is an isometry, $\beta = \phi \circ \gamma$ implies that
$$
\langle \beta'(0),
  \beta'(0) \rangle_{T_{\phi(\mathbf{x})}\mathbb{R}^d} = \langle
  \gamma'(0), \gamma'(0) \rangle_{T_{\mathbf{x}}D^m} = 1,
$$ 
so that 
$$
r({\mathbf x},{\mathbf y}) = d({\mathbf x},{\mathbf y}) = r_0.
$$
\end{proof}

\begin{lemma}
Let $\mathbf{x}, \mathbf{z}, \mathbf{y} \in D^m$ with
$r(\mathbf{x},\mathbf{z}) = d(\mathbf{x},\mathbf{z})$ and
$r(\mathbf{y},\mathbf{z}) = d(\mathbf{y},\mathbf{z})$. Then,
$r(\mathbf{x},\mathbf{y}) = d(\mathbf{x},\mathbf{y})$.
\label{lem:trans}
\end{lemma}

\begin{proof}
  We will identify the tangent spaces $T_{\mathbf{z}}D^m$ and
  $T_{\phi(\mathbf{z})}\mathbb{R}^d$ with $\mathbb{R}^m$ and
  $\mathbb{R}^d$ respectively by the standard exponential map. Since
  $r(\mathbf{x},\mathbf{z}) = d(\mathbf{x},\mathbf{z})$,
  Lemma~\ref{lem:flat} implies that $\beta_1 = \phi  \circ  \gamma_1$
  where
  $$ \gamma_1(t) = \frac{\|\mathbf{z} - \mathbf{x}\|-t}{\|\mathbf{z} -
    \mathbf{x}\|} \mathbf{x} + \frac{t}{\|\mathbf{z} - \mathbf{x}\|}
  \mathbf{z},
  $$ and
  $$ \beta_1(t) = \frac{\|\phi(\mathbf{z}) -
    \phi(\mathbf{x})\|-t}{\|\phi(\mathbf{z}) - \phi(\mathbf{x})\|}
  \phi(\mathbf{x}) + \frac{t}{\|\phi(\mathbf{z}) - \phi(\mathbf{x})\|}
  \phi(\mathbf{z}).
  $$ Similarly, $\beta_2 = \phi \circ \gamma_2$ where
  $$ \gamma_2(t) = \frac{\|\mathbf{z} - \mathbf{y}\|-t}{\|\mathbf{z} -
    \mathbf{y}\|} \mathbf{y} + \frac{t}{\|\mathbf{z} - \mathbf{y}\|}
  \mathbf{z},
  $$ and
  $$ \beta_2(t) = \frac{\|\phi(\mathbf{z}) -
    \phi(\mathbf{y})\|-t}{\|\phi(\mathbf{z}) - \phi(\mathbf{y})\|}
  \phi(\mathbf{y}) + \frac{t}{\|\phi(\mathbf{z}) - \phi(\mathbf{y})\|}
  \phi(\mathbf{z}).
  $$ Since $\phi$ is an isometric immersion, 
$$\langle \beta_1',
  \beta_2' \rangle_{T_{\phi(\mathbf{z})}\mathbb{R}^d} = \langle
  \gamma_1', \gamma_2' \rangle_{T_{\mathbf{z}}D^m}.
  $$ 
  This along with $\|\mathbf{z} - \mathbf{x}\| = \|\phi(\mathbf{z})
  - \phi(\mathbf{x})\|$ and $\|\mathbf{z} - \mathbf{y}\| =
  \|\phi(\mathbf{z}) - \phi(\mathbf{y})\|$, and using the
  identification of the tangent spaces with $\mathbb{R}^m$ and
  $\mathbb{R}^d$ yields
  $$ \langle \phi(\mathbf{x}) - \phi(\mathbf{z}), \phi(\mathbf{y}) -
  \phi(\mathbf{z}) \rangle_n = \langle \mathbf{x} - \mathbf{z},
  \mathbf{y} - \mathbf{z} \rangle_m
  $$ Therefore,
\begin{eqnarray}
  d^2(\mathbf{x},\mathbf{y}) & = & \langle \phi(\mathbf{x}) -
  \phi(\mathbf{y}), \phi(\mathbf{x}) - \phi(\mathbf{y}) \rangle_n
  \nonumber \\ & = & d^2(\mathbf{x},\mathbf{z}) +
  d^2(\mathbf{y},\mathbf{z}) - 2 \langle \phi(\mathbf{x}) -
  \phi(\mathbf{z}), \phi(\mathbf{y}) - \phi(\mathbf{z}) \rangle_n
  \nonumber \\ & = & r^2(\mathbf{x},\mathbf{z}) +
  r^2(\mathbf{y},\mathbf{z}) - 2 \langle \mathbf{x} - \mathbf{z},
  \mathbf{y} - \mathbf{z} \rangle_m \nonumber \\ & = &
  r^2(\mathbf{x},\mathbf{y}) \nonumber
\end{eqnarray}

\end{proof}

We now have all the results we need to prove
Theorem~\ref{thm:crumple}. The proof is as follows:

\begin{proof}

We will first show that for every $\mathbf{x} \in {\cal U}$, where
${\cal U}$ is as defined in Lemma~\ref{lem:dense}, there exists a
$\mathbf{y} \in \partial D^m$ such that $r(\mathbf{x},\mathbf{y})
= d(\mathbf{x},\mathbf{y})$.  We will prove this proposition by
induction.

Lemma~\ref{lem:baby_thm} implies that this statement is true if
$\mathbf{x} \in {\cal O}_0$.  Note that the statement is trivially true
for all $\mathbf{x} \in {\cal O}_p$ if ${\cal O}_p = \emptyset$. We
will assume that this statement is true for $\mathbf{x} \in {\cal
O}_j$ for $j = 0,1,2,\ldots,l-1$. Let $\mathbf{x} \in {\cal
O}_l$. Observation~\ref{obs:maximal} implies that either
\begin{enumerate}

\item The maximal geodesic starting at $\mathbf{x}$ ends on the
boundary in which case Lemma~\ref{lem:flat} proves the proposition, or

\item There exists a $\mathbf{z}^* = \gamma(b^*)$, such that the
geodesic cannot be extended in a leaf of $\Delta$ beyond
$\mathbf{z}^*$. In this case, Observation~\ref{obs:straight} implies
that $r({\mathbf x},{\mathbf z}^*) = d({\mathbf x},{\mathbf z}^*)$.

 Let $\epsilon_n : n = 1,2,\ldots $ be a decreasing
sequence with $\lim_n \epsilon_n = 0$. By Lemma~\ref{lem:baby_thm},
there exists a sequence $\mathbf{z}_n \in {\cal U}, n = 1, 2, \ldots$
with $d(\mathbf{z}_n,\mathbf{z}^*) < \epsilon_n$ and $\mathbf{z}_n \in
{\cal O}_j$ for $j = 1,2,\ldots,l-1$. By the hypothesis, there
exists a $\mathbf{y}_n \in \partial D^m$ such that
$r(\mathbf{z}_n,\mathbf{y}_n) = d(\mathbf{z}_n,\mathbf{y}_n)$.  This
procedure defines a sequence $\mathbf{y}_n \in \partial D^m$. Since
$\partial D^m$ is compact, there exists an accumulation point
$\mathbf{y}$ and a subsequence $\mathbf{y}_{n_q}$ with $n_q
\rightarrow \infty$ as $q \rightarrow \infty$ such that
$$
\lim_{q \rightarrow \infty} d(\mathbf{y},\mathbf{y}_{n_q}) = 0.
$$
It follows from the continuity of the functions
$d(\mathbf{x},\mathbf{y})$ and $r(\mathbf{x},\mathbf{y})$ with respect
to both the arguments and the fact 
$$
\lim_{q \rightarrow \infty} d(\mathbf{z}^*,\mathbf{z}_{n_q}) \leq
\lim_{q \rightarrow \infty} \epsilon_{n_q} = 0,
$$
that $d(\mathbf{z}^*,\mathbf{y}) = r(\mathbf{z}^*,\mathbf{y})$. We saw
above that $d({\mathbf x},{\mathbf z}^*) = r({\mathbf x},{\mathbf
z}^*)$. Lemma~\ref{lem:trans} now implies that
$d(\mathbf{x},\mathbf{y}) = r(\mathbf{x},\mathbf{y})$.
\end{enumerate}
This proves the proposition. 

The proof of the theorem follows from noting that since ${\cal U}$ is
dense in $D^m$, there is a sequence $\mathbf{x}_n \in {\cal U}$ such
that
$$
\lim_{n \rightarrow \infty} \mathbf{x}_n = \mathbf{x}^*,
$$ 
where $\mathbf{x}^* = \mathbf{0} \in D^m \subset \mathbb{R}^m$. By
the proposition, there exists a sequence $\mathbf{y}_n \in \partial
D^m$ with $d(\mathbf{x}_n,\mathbf{y}_n) =
r(\mathbf{x}_n,\mathbf{y}_n)$. Using the compactness of $\partial D^m$
to extract a subsequence that converges and using the continuity of
$r(.,.)$ and $d(.,.)$, it follows that there exists a $\mathbf{y}^*
\in \partial D^m$ such that $r(\mathbf{x}^*,\mathbf{y}^*) =
d(\mathbf{x}^*,\mathbf{y}^*) = 1$. Therefore,
$$
\rho(\phi) =  \sup_{\mathbf{x},\mathbf{y} \in D^m}r({\mathbf
  x},\mathbf{y}) \geq 1.
$$
\end{proof}

\section{Discussion} \label{sec:conclusions}

The theorem proved above sheds light on the global consequences of
requiring that an embedding be isometric.  Though the isometry
condition is a local one, it leads to a global constraint on the
minimal span of the embedding.  The interest of this theorem is in
identifying more general constraints of this nature.  In this section
we discuss possible generalizations and limitations of the theorem.

The mere existence of global geometric constraints arising from local
ones is not surprising.  For example if our $m$-dimensional manifold
is replaced by a $d$-dimensional solid, all the isometric embeddings
are related by rigid motions. In particular, every geodesic in the
manifold is then a geodesic in the embedding space.  The interest of
the embeddings studied above is that a global constraint exists
despite substantial allowed deformations of the manifold.  Each point
has many possibilities for bending deformation.  A given straight line
in the manifold may bend and twist in many directions.  Yet the
constraints among these bending modes due to the isometry condition
are sufficient to force the existence of at least one line through any
given point in the manifold that is also a geodesic in the embedding
space and which runs all the way to the boundary. This gives a certain
``rigidity'' to the sheet in the sense that it cannot be deformed
isometrically into an arbitrarily small sphere.

Note that this ``rigidity'' of an $m$-sheet in ${\mathbb R}^d$ for $m+1
\leq d \leq 2m-1$ is a global phenomenon in the following sense. For
the confinement of an $m$-sheet, we can always find a smooth ``local''
isometry, that is, given any small ball $B_r^d \subset {\mathbb R}^d$,
and a point $p$ in the $m$-sheet, we can find an open neighborhood
${\cal U}$ of $p$ such that there exists a smooth isometry $\psi :
{\cal U} \rightarrow B_r^d$ provided $d > m$. A consequence of the
fact that there exist such ``local'' isometries is that this ``global
rigidity'' may be easily compromised.  Though a smooth two-dimensional
sheet may not be embedded in a small sphere, a sheet with creases or
folds may be readily embedded.  To remove the rigidity, it suffices to
violate the smoothness requirement on a set of measure zero.  For a
2-sheet in 3-space, violation of smoothness at isolated points is not
sufficient; though violation on a union of one-dimensional manifolds
(the folds) is sufficient. An interesting question is the nature of
the minimal set on which we will have to violate the smoothness
requirement in order to embed an $m$-sheet in an arbitrarily small
$d$-sphere where $d < 2m$.

One natural way to weaken the isometry constraint is to treat the
manifold as an elastic object, with a local energy quadratic in the
strain that measures the departure from isometry.  If such an object
is forced into a small sphere, what shape minimizes this energy?
Simple examples suggest that, in the limit that the thickness of the
sheet goes to zero, the least costly form of deformation is to
concentrate all the strain in a small subset of the entire sheet --
the {\em singular set}, rather than deforming smoothly in which case
the strain is distributed over the entire sheet.  The energy of the
smooth deformation is typically much larger than the energy of the
singular deformation for embeddings with $d < 2m$.  In a real elastic
2-sheet embedded in 3 dimensions, this singular deformation is limited
by the finite thickness of the sheet.  However, the width of the
singular region scales as a fractional power of the
thickness~\cite{wit.sci,alex.prop}, leading to an intriguing
boundary-layer behavior.  The properties of these boundaries or ridges
of elastic 2-sheets confined in 3 dimensions have been recently
explored by energy-balance arguments, numerical calculations and
classical boundary-layer analysis~\cite{Kramer.PRL,alex,alex.prop}.

The singular regions anticipated in the confinement of elastic
manifolds resemble the topological defects seen in various condensed
matter systems \cite{mermin}. A classical example of a topological
singularity is the whorl singularity that arises if one attempts to
put a smooth non-zero vector field on a sphere. A well known result
from topology, analogous to our result in Theorem~\ref{thm:crumple},
is that there is no smooth everywhere non-zero vector field on a
sphere \cite{alg_top}.  Note however, that there do exist smooth,
everywhere nonzero, vector fields on any proper subset of the sphere.
Consequently, there exist everywhere non-zero vector fields on a
sphere that are smooth everywhere except at a single point, which is
the singular set in this case. This is very similar to the
singularities in a crumpled sheet, where, as discussed above, one has
to violate the smoothness requirement only on a small subset of the
entire sheet.  These topological singularities are explored in
homotopy theory, where one studies the mappings of a group into
another group. The topological singularities arise from the group
structure of the groups in consideration in contrast to our case where
we map a manifold into a manifold, and the singularities arise from
the presumed metric properties of the manifolds.

One may ask what useful knowledge emerges from our high-dimensional
analysis.  The specialization of our theorem to three-dimensional
embedding leads only to the well-known properties of developable
surfaces, as noted in the Introduction.  Embedding in spaces of more
than three dimensions has no obvious realization.  Yet physical
phenomena often find natural expressions in terms of high-dimensional
spaces.  Quasi-crystals may be elegantly described as
three-dimensional slices in a six-dimensional
crystal~\cite{quasicrystals}.  String theory and its recent
generalizations~\cite{str_thry} describe elementary particles and
relativistic interfaces as manifolds embedded in high-dimensional
spaces.  Finally, the dynamics of a material object is conventionally
described by embedding the three-dimensional array of particle labels
into the six-dimensional space of co-ordinates and momenta. The
constraints explored here may have implications for these cases.

In conclusion, we will point out some of the avenues for future
work. One class of interesting questions is about extending our
result. Fig.~\ref{fig:tooflat} strongly suggests that the following
result is true for the case of a 2-sheet embedded in 3 dimensions --
If $k$ is the index of relative nullity at a point $p$ in the sheet,
there is a $k$ simplex $S$ with its $k+1$ vertices on the boundary of
the sheet, such that $S$ contains $p$ and $S$ is totally geodesic in
the sheet as well as in the embedding space. This is true for almost
every point in the sheet. The only points where this doesn't hold for
the embedding in Fig.~\ref{fig:tooflat} is at three isolated points on
the boundary of the sheet, which have $k = 1$, but have no straight
lines of finite length containing them that are flat in the
embedding. Every other point in one of the regions with one curved
direction lies on a generator, which is a straight segment of finite
length with end points on the boundary. Also, every point in the
region with two flat directions, lies in the inner triangle which is a
$2$-simplex with three vertices on the boundary. This is certainly
more general than our result, which only says that there is a line
segment containing $p$ and a point on the boundary, that is totally
geodesic in the sheet and the embedding space. We believe such a
general result will hold, for this case of a 2-sheet in 3 dimensions
and also for the general case of embedding an $m$-sheet in
$d$-dimensions and we will investigate this question further in the
future.

Another interesting question is generalizing our results to the
situation where the sheets and the embedding space are not
intrinsically flat, Our theorem is restricted to flat manifolds mapped
into flat embedding spaces.  But the notion of isometric embedding
readily generalizes to curved manifolds and curved embedding spaces.
There must be restrictions on how close together the manifold points
can be brought, analogous to those of our theorem.  To formulate the
proper generalization of our theorem is a challenging task.  Hopefully
the ideas above will provide a helpful framework for this task.  

Finally, one might investigate what other kinds of local constraints
will lead to global constraints analogous to those found here. By
analogy with crumpling, we would expect that such local constraints
will give the system a global rigidity which can be destroyed by
violating the local constraints on small regions in the system. This
may prove useful in the understanding of various phenomena in
continuum physics where large scale forcing of a system can lead to
strong non-uniformities or nearly singular behavior in small localized
regions.

\renewcommand {\thesection} {Appendix}

\section{Differential Manifolds} \label{sec:manif}

We will use $\|{\mathbf x}\|$ to denote the Euclidean length of the
vector ${\mathbf x} \in {\mathbb R}^n$. A set ${\cal U} \in {\mathbb
  R}^n$ is {\em open}, if every point ${\mathbf p} \in {\cal U}$ is an
interior point, that is there exists an $r > 0$ such that
$B_{r}({\mathbf p}) = \{{\mathbf q} \in {\mathbb R}^n | \|{\mathbf p}
- {\mathbf q}\| < r \} \subseteq {\cal U}$. In what follows, by the
neighborhood of a point ${\mathbf p} \in {\mathbb R}^n$, we will mean
an open set ${\cal U}$ containing ${\mathbf p}$.

A (smooth) $n$-dimensional manifold is a set $M$ with a collection of
sets $O_{\alpha}, \alpha \in S$ where $S$ is an index set such that
\begin{enumerate}
\item The collection $\{O_{\alpha}\}, \alpha \in S$ covers $M$, that
  is
$$
M \subseteq \bigcup_{\alpha \in S} O_{\alpha}.
$$
\item For every $\alpha \in S$, there is a one-to-one, onto map
  $\psi_{\alpha} : O_{\alpha} \rightarrow {\cal U}_{\alpha}$ where
  ${\cal U}_{\alpha}$ is an open subset of ${\mathbb R}^n$.
\item if $O_{\alpha} \cap O_{\beta} \neq \emptyset$, the map 
  $$ \psi_{\beta} \circ \psi^{-1}_{\alpha} : {\cal U} \rightarrow {\mathbb
    R}^n
  $$ is smooth ($C^{\infty}$) where ${\cal U} =
  \psi_{\alpha}(O_{\alpha} \cap O_{\beta})$.
\end{enumerate}
In the rest of this appendix, we will assume that $M$ is a
$n$-dimensional manifold. The definition above can be paraphrased as
follows: Every point in $M$ has a neighborhood that looks like (is in
a one-to-one correspondence) with an open subset of ${\mathbb R}^n$.
This correspondence gives a system of co-ordinates on this
neighborhood. If there is more than one way to put co-ordinates on a
neighborhood of a point, the transformations between the various sets
of co-ordinates are smooth.

The set $O_{\alpha}$ is called a co-ordinate patch and the function
$\psi_{\alpha}$ is a co-ordinate system. The co-ordinates enable us to
go back and forth between Euclidean spaces and general manifolds and
this allows one to extend notions like continuity and smoothness which
are defined for functions between Euclidean spaces to the case of
functions between manifolds. For instance, given a function $f : M
\rightarrow {\mathbb R}^m$ and a point $p \in M$, we can choose a
co-ordinate patch $O_{\alpha}$ containing $p$ and use the associated
co-ordinate system $\psi_{\alpha}$ to define a function $f \circ
\psi^{-1}_{\alpha} : {\cal U}_{\alpha} \rightarrow {\mathbb R}^m$,
where ${\cal U}_{\alpha}$ is an open subset of ${\mathbb R}^n$ as
defined above. This procedure enables us to extend notions of
differentiability and smoothness to function $f : M \rightarrow
{\mathbb R}^m$.  We will say that $f : M \rightarrow {\mathbb R}^m$ is
differentiable (smooth) at a point $p$ if $f \circ \psi_{\alpha}^{-1} :
{\cal U}_{\alpha} \rightarrow {\mathbb R}^m$ is differentiable
(smooth) at $\psi_{\alpha}(p)$.  Likewise, a function $ g : {\mathbb
  R}^m \rightarrow M$ is differentiable (smooth) at a point ${\mathbf
  q} \in {\mathbb R}^m$, if $\psi_{\alpha} \circ f : {\mathbb R}^m
\rightarrow {\mathbb R}^n$ is smooth at ${\mathbf q}$, where
$\psi_{\alpha}$ is a co-ordinate system on a patch $O_{\alpha}$
containing $f({\mathbf q})$.

A derivation $\mathrm{v} : C^{\infty}(M) \rightarrow {\mathbb R}$ is a
map from the set of all smooth real function on $M$, to ${\mathbb R}$
that is linear $\mathrm{v}(\alpha f + \beta g) = \alpha \mathrm{v}(f)
= \beta \mathrm{v}(g)$ and satisfies the Leibniz rule $\mathrm{v}(fg)
= \mathrm{v}(f)g + \mathrm{v}(g)f$, for all $\alpha,\beta \in \
{\mathbb R}$ and all smooth functions $f,g$. The set of all the
derivations at a point $p \in M$ is a $n$ dimensional vector space
called the tangent space at $p$, and is denoted by $T_pM$. The
elements of this vector space are called (tangent) vectors at the {\em
base point} $p$.

The union of all the tangent spaces as the base point $p$ ranges over
all the points in the manifold is called the {\em tangent bundle} and
is denoted by $TM$, that is
$$
TM = \bigcup_{p \in M} T_pM.
$$ The tangent bundle is equipped with a natural projection operator
$\pi : TM \rightarrow M$, where $\pi(X) = p$ if $X \in T_pM$, that is,
$\pi$ gives the base point corresponding to a given tangent vector. We
will use upper case letters $W,X,Y,Z,\ldots$ to denote vectors (at a
given base point $p$) and vector fields, that is functions $X: M
\rightarrow TM$ such that $X(p) \in T_p(M)$.

A differentiable function $\gamma : [a,b] \rightarrow M$ where $[a,b]
\subseteq {\mathbb R}$ is called a (parameterized) curve.  Given a
curve $\gamma : [a,b] \rightarrow M$, if $p \in M$ is such that
$\gamma(c) = p$ for some $c \in (a,b)$, we can define a derivation at
$p$ by $\mathrm{v}(f) = (f \circ \gamma)'(c)$. We will call the vector
$\mathrm{v}$ defined by this procedure the tangent vector to the curve
at $p$ and denote it by $\gamma'(t)$ or $d\gamma/dt$.

If $p$ is any point in $M$ and $O_{\alpha}$ is a co-ordinate patch
containing $p$, we can define a basis for the tangent space $T_pM$
using the derivations given by the curves $\gamma_i, 1\leq i \leq n$
with $\gamma_i(t) = \psi_{\alpha}^{-1} (\psi(p) + t {\mathbf e}_i)$,
where ${\mathbf e}_1,{\mathbf e}_2,\ldots,{\mathbf e}_n$ is the
standard basis of ${\mathbb R}^n$. (The curves $\gamma_i$ are obtained
by first choosing a co-ordinate system at $p$, starting at $p$,
keeping $n-1$ co-ordinates fixed and varying just one co-ordinate.)
We will say that a vector field $X$ is smooth at $p$ if its
components in this basis are smooth functions in the co-ordinate
system $\psi_{\alpha}$ in a neighborhood of $\psi_{\alpha}(p)$.
Different sets of co-ordinates will give different bases and differing
values for the components, but they give the same notion of smooth
vector fields since the co-ordinates are related by smooth
transformations.  However, we cannot define a derivative operator on
vector fields because there does not exist a preferred set of
co-ordinates and in general the relation between the basis of $T_pM$
and $T_qM$ for two distinct points $p$ and $q$ is not an intrinsic
property of the manifold $M$, but it depends on the co-ordinates
chosen near $p$ and $q$. A derivative operator or a connection
$\nabla$ is an additional structure that is imposed on the manifold
(more properly, the tangent bundle $TM$) that enables one to compare
vectors (or generally tensors) at two distinct base points in the
manifold.  If $\gamma : [a,b] \rightarrow M$ is a curve and $Y$ is a
smooth vector field, then the derivative of $Y$ along the curve
$\gamma$ at a point $p$ is denoted by $\nabla_X Y$, where $X$ is the
tangent to $\gamma$ at $p$.  A connection $\nabla$ should be
compatible with the derivations discussed above in the sense that
$X(f) = \nabla_X f$ for all vector fields $X$ and all smooth functions
$f : M \rightarrow {\mathbb R}$. The connection should also satisfy
the Leibniz rule for all products of tensors on $M$ and should be
torsion--free~\cite{waldbook}. 

If a manifold has a local measure of distance, there exists a function
$|.|_p : T_p M \rightarrow {\mathbb R}$, such that $|\gamma'(c)|_p$ is
the speed of a curve $\gamma(t)$ at $p = \gamma(c)$. A Riemannian
manifold $M$ is one where this function is given by the norm
corresponding to a non-degenerate, positive-definite inner product
$\langle ., .  \rangle_p : T_pM \times T_pM \rightarrow {\mathbb R}$ ,
{\em i.e.}, $|X|_p = \sqrt{\langle X, X \rangle_p}$ for all $X \in
T_pM$. The inner product $\langle ., .  \rangle_p : T_pM \times T_pM
\rightarrow {\mathbb R}$ is called the metric.

On a Riemannian manifold, there is a unique connection $\nabla$ such
that $ \nabla_X (\langle Y, Z \rangle_p) = \langle Y, \nabla_X Z
\rangle_p + \langle \nabla_X Y, Z \rangle_p$ for all vector fields
$X,Y,Z$ defined in a neighborhood of $p$. This is called the
Levi-Civita connection and unless otherwise noted, this is the
connection that we will use.

We will now consider mappings of a manifold $M$ into another manifold
$\tilde M$.  Such a mapping $\phi: M \rightarrow \tilde M$ is {\it
  smooth} if its representation in terms of co-ordinates on $M$ and
$\tilde{M}$ is smooth. If $\phi : M \rightarrow \tilde{M}$ is a smooth
mapping, any smooth function $\tilde{f} : \tilde{M} \rightarrow
{\mathbb R}$ can be pulled back to yield a smooth function $f : M
\rightarrow {\mathbb R}$ by $f(p) = \tilde{f}(\phi(p))$. If
$\mathrm{v}$ is a derivation at $p$, it is easily verified that
$\tilde{\mathrm{v}}$ defined by $\tilde{\mathrm{v}}(\tilde{f}) =
\mathrm{v}(\tilde{f}\circ\phi)$ is a derivation on
$C^{\infty}(\tilde{M})$. The map $\phi$ therefore induces a map
between the tangent spaces $T_p(M)$ and $T_{\phi(p)}\tilde{M}$ given
by $\mathrm{v} \mapsto \tilde\mathrm{v}$, and this map is
conventionally denoted by $\phi_*$.

A smooth mapping $\phi:M \rightarrow \tilde{M}$ is an {\em immersion}
if $\phi_* : T_p M \rightarrow T_{\phi(p)} \tilde{M}$ is one-to-one.
This definition can be paraphrased as follows -- $\phi$ is an
immersion only if it is locally one-to-one, that is given any point
$p$ in $M$, there is an open set $O$ containing $p$ such that the
image $\phi(O)$ does not intersect itself in $\tilde{M}$ or fold back
on itself so that it comes arbitrarily close to self-intersection.

For a complete and mathematically rigorous discussion of differential
geometry of manifolds see~\cite{manifolds}.

\renewcommand{\thesection} {}

\section{Acknowledgments} 

This work was supported by the National Science Foundation under
awards number DMR 9528957, DMR 9975533 and DMR 9808595. S.C.V. was
also supported by a Research fellowship from the Alfred P. Sloan
Jr. Foundation.

\end {document}